\documentclass[a4paper,11pt]{article}

\usepackage[hidelinks]{hyperref}
\usepackage[table]{xcolor}
\hypersetup{
    colorlinks,
    linkcolor={blue!40!black},
    citecolor={red!40!black},
    urlcolor={green!20!black}
}

\usepackage{graphicx}
\usepackage{hhline}
\usepackage{subfig}

\usepackage{ulem}
\usepackage{colortbl}
\usepackage{multirow}
\usepackage{tikz}

\usepackage{url}

\usepackage{xspace}

\usepackage{amssymb}
\usepackage{amsmath}
\usepackage{enumerate}
\usepackage{algorithm}
\usepackage{algpseudocode}

\usepackage{amsthm}

\usepackage{gensymb}

\usepackage{hhline}

\usepackage{makecell}

\usepackage{tabularx, booktabs}
\usepackage{array}

\makeatletter
\newcommand{\thickhline}{%
    \noalign {\ifnum 0=`}\fi \hrule height 1.5pt
    \futurelet \reserved@a \@xhline
}
\newcolumntype{"}{@{\hskip\tabcolsep\vrule width 1pt\hskip\tabcolsep}}
\makeatother

\newcolumntype{?}{!{\vrule width 1pt}}

\newtheorem{definition}{Definition}

\newcommand{\interface}[1]{\textsc{{#1}}}
\newcommand{\msgtag}[1]{\textsc{{#1}}}
\newcommand{\angular}[1]{\ensuremath{\langle}#1\ensuremath{\rangle}}
\newcommand{\msg}[1]{\angular{\ensuremath{#1}}}

\newcommand{\MSGPREPARE}{\textsc{prepare}}
\newcommand{\MSGCOMMIT}{\textsc{commit}}

\newcommand{\MSGREPLY}{\textsc{reply}}

\newcommand{\PCBFT}{\textsf{XPaxos}\xspace}
\newcommand{\ndR}{replicate}

\newcommand{\omitit}[1]{}

\newtheorem{lemma}{Lemma}

\newcommand{\req}{\ensuremath{req}\xspace}

\newcommand{\vs}{\ensuremath{sg}\xspace}
\newcommand{\sg}{\ensuremath{sg}\xspace}

\newcommand{\rep}{\ensuremath{rep}\xspace}
\newcommand{\sn}{\ensuremath{sn}\xspace}

\usepackage{mathtools}

\algnewcommand{\LineComment}[1]{\Statex \hskip\ALG@thistlm \(\triangleright\) #1}

\newcommand{\cmt}[1]{\hfill /* #1 */}

\newcommand{\process}{machine\xspace}
\newcommand{\processes}{machines\xspace}
\newcommand{\replica}{replica\xspace}
\newcommand{\replicas}{replicas\xspace}
\newcommand{\Replicas}{\Pi}
\newcommand{\clients}{C}

\newcommand{\pcorrect}{p_{\textit{correct}}}
\newcommand{\pcrash}{p_{\textit{crash}}}
\newcommand{\pnoncrash}{p_{\textit{non-crash}}}
\newcommand{\pbenign}{p_{\textit{benign}}}

\newcommand{\pnormal}{p_{\textit{available}}}

\newcommand{\timeout}{\Delta}
\newcommand{\psynchrony}{p_{\textit{synchrony}}}
\newcommand{\ninessynchrony}{9_{\textit{synchrony}}}

\newcommand{\ninesOf}{\textit{9of}}
\newcommand{\ninesbenign}{9_{\textit{benign}}}
\newcommand{\ninesofC}{\textit{9ofC}}  
\newcommand{\ninesofA}{\textit{9ofA}} 
\newcommand{\ninesnormal}{9_{\textit{available}}}
\newcommand{\ninescorrect}{9_{\textit{correct}}}

\usepackage{geometry}

\usepackage[titletoc,title]{appendix}

\begin{document}
	
	\title{XFT: Practical Fault Tolerance Beyond Crashes}
\author{Shengyun Liu \\ NUDT\thanks{Work done while being a PhD student at EURECOM.}\\\emph{lius@eurecom.fr} \and Paolo Viotti\\ EURECOM\\\emph{paolo.viotti@eurecom.fr} \and Christian Cachin\\IBM Research - Zurich\\ \emph{cca@zurich.ibm.com} \and Vivien Qu\'ema\\INP Grenoble\\\emph{vivien.quema@imag.fr} \and Marko Vukoli\'c\\IBM Research - Zurich \\ \emph{mvu@zurich.ibm.com}}

	\date{}	
	\maketitle
	
	\begin{abstract}
		Despite years of intensive research, 
		Byzantine fault-tolerant (BFT) systems have not yet been adopted in practice. This is due to additional cost of BFT in terms of resources, protocol complexity and performance, compared with crash fault-tolerance (CFT). 
		This overhead of BFT comes from the assumption of a powerful adversary that can fully \emph{control} not only the Byzantine faulty machines, but \emph{at the same time} also the message delivery schedule across the \emph{entire} network, effectively inducing communication asynchrony and partitioning otherwise correct machines at will. To many practitioners, however, such strong attacks appear irrelevant.
		
		In this paper, we introduce \emph{cross fault tolerance} or \emph{XFT}, a novel approach to building reliable and secure distributed systems and apply it to the classical state-machine replication (SMR) problem. In short, an XFT SMR protocol provides the reliability guarantees of widely used asynchronous CFT SMR protocols such as Paxos and Raft, but also tolerates Byzantine faults in combination with network asynchrony, as long as a majority of replicas are correct and communicate synchronously. This allows the development of XFT systems at the price of CFT (already paid for in practice), yet with strictly stronger resilience than CFT --- sometimes even stronger than BFT itself.
		
		As a showcase for XFT, we present 
		\PCBFT, the first XFT SMR protocol, and deploy it in a geo-replicated setting. 
		Although it offers much stronger resilience than CFT SMR at no extra resource cost, the performance of \PCBFT  matches that of the state-of-the-art CFT protocols.  

	\end{abstract}
	
	\section{Introduction}
	
	Tolerance to any kind of service disruption, whether caused by a simple hardware fault or by a large-scale disaster, is key for the survival of modern distributed systems.  Cloud-scale applications must be inherently resilient, as any outage has direct implications on the business behind them~\cite{Krishnan:2012}. 
	
	Modern production systems (e.g., \cite{Corbett:2012:SGG:2387880.2387905, Calder:2011:WAS:2043556.2043571}) increase the number of \emph{nines of reliability}\footnote{As an illustration, five nines reliability means that a system is up and correctly running at least 99.999\% of the time.  In other words, malfunction is limited to one hour every 10 years on average.} by employing sophisticated distributed protocols that tolerate \emph{crash} machine faults as well as \emph{network faults}, such as network partitions or asynchrony, which reflect the inability of otherwise \emph{correct} machines to communicate among each other in a timely manner. At the heart of these systems typically lies a crash fault-tolerant (CFT) consensus-based \emph{state-machine replication} (SMR) primitive~\cite{Schneider90,Chandra:2007:PML:1281100.1281103}.
	
	These systems cannot deal with \emph{non-crash} (or \emph{Byzantine} \cite{Lamport:1982:BGP:357172.357176}) faults, which include not only malicious, adversarial behavior, but also arise from errors in the hardware, stale or corrupted data from storage systems, memory errors caused by physical effects, bugs in software, hardware faults due to ever smaller 
 circuits, and human mistakes that cause state corruptions and data loss.  However, such problems do occur in practice --- each of these faults has a public record of taking down major production systems and corrupting their service \cite{ASC,Bailis14}.
	
	Despite more than 30 years of intensive research since the seminal work of  Lamport, Shostak and Pease~\cite{Lamport:1982:BGP:357172.357176}, no \textit{practical} answer to tolerating non-crash faults has emerged so far. In particular, asynchronous Byzantine fault-tolerance (BFT), which promises to resolve this problem~\cite{Castro:2002:PBF}, has not lived up to this expectation, largely because of its extra cost compared with CFT.  Namely, asynchronous (that is,  ``eventually synchronous'' \cite{DLS}) BFT SMR must use at least $3t+1$ replicas to tolerate $t$ non-crash faults \cite{BrachaT85} instead of only $2t+1$ replicas for CFT, as used by Paxos \cite{Lamport:1998:PP:279227.279229} or Raft \cite{RAFT}, for example.
	
	The overhead of asynchronous BFT is due to the extraordinary power given to the adversary, which may control both the Byzantine faulty machines \emph{and} the \emph{entire network} in a coordinated way. In particular, the classical BFT adversary can partition \emph{any number} of otherwise correct machines at will. In line with observations by practitioners~\cite{bftw3}, we claim that this adversary model is actually too strong for the phenomena observed in deployed systems. For instance, accidental non-crash faults usually do not lead to network partitions. Even malicious non-crash faults rarely cause the whole network to break down in wide-area networks and geo-replicated systems. The proverbial all-powerful attacker as a common source behind those faults is a popular and powerful simplification used for the design phase, but it has not seen equivalent proliferation in practice. 
	
	In this paper, we introduce \emph{XFT} (short for \emph{cross fault tolerance}), a novel approach to building efficient resilient distributed systems that tolerate both non-crash (Byzantine) faults and network faults (asynchrony). In short, XFT allows building resilient systems that
	\begin{itemize}
		\item  do not use extra resources (replicas) compared with asynchronous CFT; 
		\item  preserve \emph{all} reliability guarantees of asynchronous CFT (that is, in the absence of Byzantine faults); and
		\item  provide correct service (i.e., safety and liveness \cite{Alpern:1987}) even when Byzantine faults do occur, as long as a majority of the replicas are correct and can communicate with each other synchronously (that is, when a minority of the replicas are Byzantine-faulty or partitioned because of a network fault). 
	\end{itemize}
		
	 In particular, we envision XFT for wide-area or \emph{geo-replicated} systems~\cite{Corbett:2012:SGG:2387880.2387905}, as well as for any other deployment where an adversary cannot easily coordinate enough network partitions and  Byzantine-faulty machine actions at the same time. 
	
	As a showcase for XFT, we present \PCBFT, the first state-machine replication protocol in the XFT model. \PCBFT  tolerates faults beyond crashes in an efficient and practical way, achieving much greater coverage of realistic failure scenarios than the state-of-the-art CFT SMR protocols, such as Paxos or Raft. 
	This comes without resource overhead as \PCBFT uses $2t+1$ replicas. To validate the performance of \PCBFT, we deployed it in a geo-replicated setting across Amazon EC2 datacenters worldwide. In particular, we integrated \PCBFT within Apache ZooKeeper, a prominent and widely used coordination service for cloud systems~\cite{Zookeeper}. Our evaluation on EC2 shows that \PCBFT performs almost as well in terms of throughput and latency as a WAN-optimized variant of Paxos, and significantly better than the best available BFT protocols. In our evaluation, \PCBFT even outperforms the native CFT SMR protocol built into  ZooKeeper \cite{Zab}.

	 Finally, and perhaps surprisingly, we show that XFT can offer \emph{strictly stronger} reliability guarantees than state-of-the-art BFT, for instance under the assumption that machine faults and network faults occur as independent and identically distributed random variables, for certain probabilities. To this end, we  calculate the number of nines of consistency (system safety) and availability (system liveness) of resource-optimal CFT, BFT and XFT (e.g., \PCBFT) protocols. Whereas XFT \emph{always} provides strictly stronger consistency and availability guarantees than CFT and \emph{always} strictly stronger availability guarantees than BFT, our reliability analysis shows that, in some cases, XFT also provides strictly stronger consistency guarantees than BFT.
	
	The remainder of this paper is organized as follows. In Section~\ref{sec:xpaxos_model}, we define the system model, which is then followed by the definition of the XFT model in Section~\ref{sec:xft}. In Section~\ref{sec:nutshell} and Section~\ref{sec:evaluation}, we present \PCBFT and its evaluation in the geo-replicated context, respectively. Section~\ref{sec:nines}  provides simplified reliability analysis comparing XFT with CFT and BFT. We overview related work and conclude in Section~\ref{sec:conclusion}. The full pseudocode and correctness proof of \PCBFT is given in Appendix~\ref{apd:pseudocode} and~\ref{apd:proof}. 
	
	\section{System model}
	\label{sec:xpaxos_model}
	
	\newtheorem{theorem}{\textbf{Theorem}}
	
	\noindent\textbf{Machines.} We consider a message-passing distributed system containing a set $\Replicas$ of $n = |\Replicas|$ \emph{\processes}, also called \emph{\replicas}. Additionally, there is a separate set $\clients$ of \emph{client} \processes.
	
	Clients and replicas may suffer from Byzantine \emph{faults}: we distinguish between \emph{crash} faults, where a \process simply stops all computation and communication, and \emph{non-crash} faults, where a \process acts arbitrarily, but cannot break cryptographic primitives we use (cryptographic hashes, MACs, message digests and digital signatures). A \process that is not faulty is called \emph{correct}. We say a \process is \emph{benign} if the \process is correct or crash-faulty. We further denote the number of replica faults at a given moment $s$ by
	
	\begin{itemize}
		\item $t_c(s)$: the number of crash-faulty \replicas, and
		\item $t_{nc}(s)$: the number of non-crash-faulty \replicas.
	\end{itemize}
	
	\noindent\textbf{Network.} Each pair of \replicas is connected with reliable point-to-point bi-directional communication channels. In addition, each client can communicate with any \replica.
	
	The system can be \emph{asynchronous} in the sense that \processes may not be able to exchange messages and obtain responses to their requests in time. In other words, \emph{network faults} are possible; we define a \emph{network fault} as the inability of some \emph{correct} \replicas to communicate with each other in a timely manner, that is, when a message exchanged between two correct \replicas cannot be delivered and processed within delay~$\timeout$, known to all \replicas. Note that $\timeout$ is a deployment specific parameter: we discuss practical choices for $\timeout$ in the context of our geo-replicated setting in Section~\ref{sec:evaluation}.  Finally, we assume an \emph{eventually synchronous} system in which, eventually, network faults do not occur~\cite{DLS}.   
	
	Note that we model an excessive processing delay as a network problem and \emph{not} as an issue related to a \process fault.  This choice is made consciously, rooted in the experience that for the general class of protocols considered in this work, a long local processing time is never an issue on correct \processes compared with network delays.
	
	To help quantify the number of network faults, we first give the definition of partitioned \replica.

	\begin{definition}[Partitioned \replica]
		\label{def:partition}
		Replica $p$ is partitioned if $p$ is \textbf{not} in the largest subset of \replicas, in which every pair of \replicas can communicate among each other within delay~$\timeout$. 
	\end{definition}

\begin{figure}[!htbp]
		\begin{center}
			\includegraphics[scale=0.60]{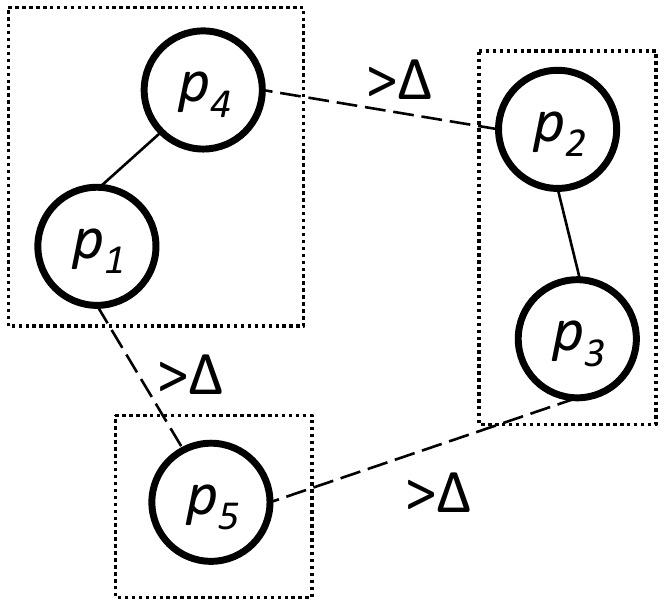}
			\caption{An illustration of partitioned \replicas : $\{p_1, p_4,p_5\}$ or $\{p_2,p_3,p_5\}$ are partitioned based on Definition~\ref{def:partition}.}
			\label{fig:partition}
		\end{center}
	\end{figure}

	If there is more than one subset with the maximum size, only one of them is recognized as the largest subset. For example in Figure~\ref{fig:partition}, the number of partitioned \replicas is 3, counting either the group of $p_1$, $p_4$ and $p_5$ or that of $p_2$, $p_3$ and $p_5$. The number of partitioned \replicas can be as much as $n-1$, which means that no two \replicas can communicate with each other within delay $\timeout$. We say \replica $p$ is \emph{synchronous} if $p$ is not partitioned. We now quantify network faults at a given moment $s$ as

	\begin{itemize}
		\item $t_p(s)$: the number of correct, but partitioned \replica{}s.
	\end{itemize}

\smallskip
	
	\noindent\textbf{Problem.} In this paper, we focus on the \emph{deterministic} state-machine replication problem (SMR) \cite{Schneider90}. In short, in SMR clients invoke requests, which are then committed by replicas. SMR ensures
	\begin{itemize}
		\item \emph{safety}, or \emph{consistency}, by (a) enforcing \emph{total order} across committed client's \emph{requests} across all correct replicas; and by (b) enforcing \emph{validity}, i.e., that a correct replica commits a request only if it was previously invoked by a client;
		\item  \emph{liveness}, or \emph{availability}, by eventually committing a request by a correct client at all correct replicas and returning an application-level reply to the client. 
	\end{itemize}
	
	\section{The XFT model}
	\label{sec:xft}
	
	This section introduces the XFT model and relates it to the established crash-fault tolerance (CFT) and Byzantine-fault tolerance (BFT) models. 
	
	\subsection{XFT in a nutshell}
	
	\begin{table*}[!htbp]
		\centering
		\small
		\renewcommand{\arraystretch}{1.5}
		\makebox[\textwidth][c]{
		\begin{tabular}{|c|c||c|c|c|}
			\cline{3-5}
			\multicolumn{2}{c|}{} & \multicolumn{3}{c|}{Maximum number of each type of \replica faults}\\
			\cline{3-5}
			\multicolumn{2}{c|}{} & non-crash faults & crash faults & partitioned \replicas\\
			\Xhline{2\arrayrulewidth}
			\multirow{2}{*}{Asynchronous CFT (e.g., Paxos \cite{lamport2001paxos})} & consistency & 0 & $n$ & $n-1$ \\
			\cline{2-5}
			& availability & 0 & \multicolumn{2}{c|}{$\lfloor \frac{n-1}{2} \rfloor$ (combined)} \\
							\Xhline{2\arrayrulewidth}
			\multirow{2}{*}{Asynchronous BFT (e.g., PBFT \cite{Castro:2002:PBF})} & consistency & $\lfloor \frac{n-1}{3} \rfloor$ & $n$ & $n-1$\\
			\cline{2-5}
			& availability & \multicolumn{3}{c|}{$\lfloor \frac{n-1}{3} \rfloor$ (combined)}
			\\				\Xhline{2\arrayrulewidth}
			\multirow{2}{*}{(Authenticated) Synchronous BFT (e.g.,  \cite{Lamport:1982:BGP:357172.357176})} & consistency & $n-1$ & $n$ & 0 \\
			\cline{2-5}
			& availability & \multicolumn{2}{c|}{$n-1$ (combined)} & 0\\
			\Xhline{2\arrayrulewidth}
			\multirow{3}{*}{XFT (e.g., \PCBFT)} & \multirow{2}{*}{consistency} & 0 & $n$ & $n-1$\\
			\cline{3-5}
			& & \multicolumn{3}{c|}{$\lfloor \frac{n-1}{2} \rfloor$ (combined)}\\
			\cline{2-5}
			& availability & \multicolumn{3}{c|}{$\lfloor \frac{n-1}{2} \rfloor$ (combined)}\\
			\hline
		\end{tabular}}
		\caption{The maximum numbers of each type of fault tolerated by representative SMR protocols. Note that XFT provides consistency in two modes, depending on the occurrence of non-crash faults.}
		\label{table:comp}
	\end{table*}
	
	Classical CFT and BFT explicitly model \process faults only. These are 	then combined with an orthogonal network fault model, either the synchronous model (where network faults in our sense are ruled out), or the asynchronous model (which includes \emph{any number} of network faults). Hence, previous work can be classified into four categories: synchronous CFT~\cite{CristianASD95,Schneider90}, asynchronous CFT \cite{Schneider90, Lamport:1998:PP:279227.279229, Oki:1988:VRN:62546.62549},
	synchronous BFT~\cite{Lamport:1982:BGP:357172.357176,dolstr83,begape89}, and asynchronous BFT~\cite{Castro:2002:PBF,Aublin:2015:NBP:2723895.2658994}.

	XFT, in contrast, redefines the boundaries between machine and network fault dimensions: XFT allows the design of reliable protocols that tolerate crash \process faults regardless of the number of network faults and that, at the same time, tolerate non-crash \process faults when the number of machines that are either faulty or partitioned is within a threshold. 
	
	To formalize XFT, we first define \emph{anarchy}, a very severe system condition with actual non-crash machine (replica) faults and plenty of faults of different kinds, as follows:
		
		\begin{definition}[Anarchy]
			\label{def:anarchy}
			The system is in anarchy at a given moment $s$ iff $t_{nc}(s)>0$ and $t_c(s)+t_{nc}(s)+t_p(s)>t$.
		\end{definition}
		
		Here, $t$ is the threshold of replica faults, such that $t\le \lfloor \frac{n-1}{2} \rfloor$. In other words, in anarchy, some replica is non-crash-faulty, and there is no correct and synchronous majority of replicas. Armed with the definition of anarchy, we can define XFT protocols for an arbitrary distributed computing problem in function of its safety property \cite{Alpern:1987}.
		
		\begin{definition}[XFT protocol]
		\label{def:xft}
			Protocol $P$ is an XFT protocol if $P$ satisfies safety in all executions in which the system is never in anarchy.
		\end{definition}
		
		Liveness of an XFT protocol will typically depend on a problem and implementation. For instance, for deterministic SMR we consider in this paper, our \PCBFT protocol eventually satisfies liveness, provided a majority of replicas is correct and synchronous. This can be shown optimal.
	
	\omitit{
	With XFT, for example, we can design a practical SMR protocol (\PCBFT, Sec.~\ref{sec:nutshell}) that uses only $n = 2t+1$ replicas, yet tolerates up to $t$ \emph{non-crash} faulty replicas and partitioned replicas, so long as the system remains out of anarchy.

	This  greatly improves the flexibility in the choice of practically relevant fault scenarios.
	In a nutshell, XFT 	assumes that \process and network faults in a distributed system arise from different and independent causes. 
}
	
	\subsection{XFT vs. CFT/BFT}
	
	 Table~\ref{table:comp} illustrates differences between XFT and CFT/BFT in terms of their consistency and availability guarantees for SMR.
	
	State-of-the-art asynchronous CFT protocols~\cite{lamport2001paxos,RAFT} guarantee consistency despite \emph{any} number of crash-faulty \replicas and \emph{any} number of partitioned \replicas. They also guarantee availability whenever a majority of replicas  ($t \le \lfloor \frac{n-1}{2} \rfloor$) are correct and synchronous. As soon as a single machine is non-crash-faulty, CFT protocols guarantee neither consistency nor availability.
	
	Optimal asynchronous BFT protocols~\cite{Castro:2002:PBF, Kotla:2009:ZSB, Aublin:2015:NBP:2723895.2658994} guarantee consistency despite any number of crash-faulty or partitioned \replicas,  with at most $t = \lfloor \frac{n-1}{3}\rfloor$ non-crash-faulty \replicas. 	They also guarantee availability with up to  $\lfloor \frac{n-1}{3} \rfloor$ combined faults, i.e., whenever more than two-thirds of replicas are correct and not partitioned.
	Note that BFT availability might be weaker than that of CFT in the absence of non-crash faults --- unlike CFT, BFT does not guarantee availability when the sum of crash-faulty and partitioned replicas is in the range $[n/3,n/2)$.
	
	Synchronous BFT protocols (e.g., \cite{Lamport:1982:BGP:357172.357176}) do not consider the existence of correct, but partitioned replicas. This makes for a very strong assumption --- and helps synchronous BFT protocols that use digital signatures for message authentication (so called \emph{authenticated} protocols) to tolerate up to $n-1$ non-crash-faulty replicas. 
	
	In contrast, XFT protocols with optimal resilience, such as our \PCBFT, guarantee consistency in two modes: \emph{(i)} without non-crash faults, despite any number of crash-faulty and partitioned replicas (i.e., just like CFT), and \emph{(ii)} with non-crash faults, whenever a majority of replicas are correct and not partitioned, i.e., provided the sum of all kinds of faults (machine or network faults) does not exceed $\lfloor \frac{n-1}{2} \rfloor$.
	Similarly, it also guarantees availability whenever a majority of replicas are correct and not partitioned.
	
	It may be tempting to view XFT as some sort of a combination of the asynchronous CFT and synchronous BFT models. However, this is misleading, as even with actual non-crash faults, XFT is incomparable to authenticated synchronous BFT. Specifically, authenticated synchronous BFT  protocols, such as the seminal Byzantine Generals protocol \cite{Lamport:1982:BGP:357172.357176}, may violate consistency with a single partitioned replica. For instance, with $n=5$  replicas and an execution in which three replicas are correct and synchronous, one replica is correct but partitioned and one replica is non-crash-faulty, the XFT model mandates that the consistency be preserved, whereas the Byzantine Generals protocol may violate consistency.\footnote{XFT is not stronger than authenticated synchronous BFT either, as the latter tolerates more \process faults in the complete absence of network faults.}
	
	Furthermore, from Table~\ref{table:comp}, it is evident that XFT offers strictly stronger guarantees than asynchronous CFT, for both availability and consistency. XFT also offers strictly stronger availability guarantees than asynchronous BFT. Finally, the consistency guarantees of XFT are incomparable to those of asynchronous BFT.  On the one hand, outside anarchy, XFT is consistent with the number of non-crash faults in the range $[n/3,n/2)$, whereas asynchronous BFT is not. On the other hand, unlike XFT, asynchronous BFT is consistent in anarchy provided the number of non-crash faults is less than $n/3$. 
	We discuss these points further in Section~\ref{sec:nines}, where we also quantify the reliability comparison between XFT and asynchronous CFT/BFT assuming the special case of independent faults.

	\subsection{Where to use XFT?}
	
	 The intuition behind XFT starts from the assumption that ``extremely bad'' system conditions, such as anarchy, are very rare, and that providing consistency guarantees in anarchy might not be worth paying the asynchronous BFT premium. 
	 
	 In practice, this assumption is plausible in many deployments. We envision XFT for use cases in which an adversary cannot easily coordinate enough network partitions and non-crash-faulty  machine actions at the same time. Some interesting candidate use cases include:
	 
	 \begin{itemize}
	 	\item \emph{Tolerating ``accidental'' non-crash faults.} In systems which are not susceptible to malicious behavior and deliberate attacks, XFT can be used to protect against ``accidental`` non-crash faults, which can be assumed to be largely independent of network faults. In such cases, XFT could be used to harden CFT systems without considerable overhead of BFT. 
	 	\item \emph{Wide-area networks and geo-replicated systems.} XFT may reveal useful even in cases where the system is susceptible to malicious non-crash faults, as long as it may be difficult or expensive for an adversary to coordinate an attack to compromise Byzantine \processes and partition sufficiently many replicas \emph{at the same time}. Particularly interesting for XFT are WAN and geo-replicated systems which often enjoy redundant communication paths and typically have a smaller surface for network-level DoS attacks (e.g., no multicast storms and flooding). 
	 	\item \emph{Blockchain.} A special case of geo-replicated systems, interesting to XFT, are blockchain systems. In a typical blockchain system, such as Bitcoin \cite{Nakamoto:Bitcoin}, participants may be financially motivated to act maliciously, yet may lack the means and capabilities to compromise the communication among (a large number of) correct participants. In this context, XFT is particularly interesting for so-called permissioned blockchains, which are based on state-machine replication rather than on Bitcoin-style proof-of-work 
	 	\cite{Vukolic15}.
	 \end{itemize} 
	 
	\section{\PCBFT Protocol}
	\label{sec:nutshell}
	
	\subsection{\PCBFT overview}
	
	\PCBFT is a novel state-machine replication (SMR) protocol designed specifically in the XFT model. \PCBFT specifically targets good performance in geo-replicated settings, which are characterized by the network being the bottleneck, with high link latency and relatively low, heterogeneous link bandwidth. 

	In a nutshell, \PCBFT consists of three main components:
	\begin{itemize}
		\item A common-case protocol, which replicates and totally orders requests across replicas. This has, roughly speaking, the message pattern and complexity of communication among replicas of state-of-the-art CFT protocols (e.g., Phase 2 of Paxos), hardened by the use of digital signatures.   
		\item A novel view-change protocol, in which the information is transferred from one view (system configuration) to another in a \emph{decentralized}, leaderless fashion.
		\item A fault detection (FD) mechanism, which can help detect, outside anarchy, non-crash faults that would leave the system in an inconsistent state in anarchy. The goal of the FD mechanism is to minimize the impact of long-lived non-crash faults (in particular ``data loss'' faults) in the system and to help detect them before they coincide with a sufficient number of crash faults and network faults to push the system into anarchy.  
	\end{itemize}
	
	\PCBFT\ is orchestrated in a sequence of \emph{views} \cite{Castro:2002:PBF}. The central idea in \PCBFT is that, during common-case operation in a given view, \PCBFT synchronously replicates clients' requests to only $t+1$ replicas, which are the members of a \emph{synchronous group} (out of $n=2t+1$ replicas in total). Each view number $i$ uniquely determines the synchronous group, $\vs_{i}$, using a mapping known to all replicas. Every synchronous group consists of one \emph{primary} and $t$ \emph{followers}, which are jointly called \emph{active replicas}. The remaining $t$ replicas in a given view are called \emph{passive} replicas; optionally, passive replicas learn the order from the active replicas using the \emph{lazy replication} approach \cite{Lazy}. A view is not changed unless there is a machine or network fault within the synchronous group. 
	
	In the common case (Section~\ref{sec:CC}), the clients send digitally signed requests to the primary, which are then replicated across $t+1$ active replicas. These $t+1$ replicas digitally sign and locally log the proofs for all replicated requests to their \emph{commit logs}. Commit logs then serve as the basis for maintaining consistency in view changes. 
	
	The view change of \PCBFT (Section~\ref{sec:VC}) reconfigures the entire synchronous group, not  only the leader. All $t+1$ active replicas of the new synchronous group $\vs_{i+1}$ try to transfer the state from the preceding views to view $i+1$. This \emph{decentralized} approach to view change stands in sharp contrast to the classical reconfiguration/view-change in CFT and BFT protocols (e.g., \cite{Lamport:1998:PP:279227.279229,Castro:2002:PBF}), in which only a single replica (the primary) leads the view change and transfers the state from previous views. This difference is crucial to maintaining consistency (i.e., total order) across \PCBFT\ views in the presence of non-crash faults (but in the absence of full anarchy).  This  novel and decentralized view-change scheme of \PCBFT guarantees that even in the presence of non-crash faults, but outside anarchy, at least one correct replica from the new synchronous group $\vs_{i+1}$ will be able to transfer the correct state from previous views, as it will be able to contact some correct replica from any old synchronous group.

	Finally, the main idea behind the FD scheme of \PCBFT is the following. In view change, 
	a non-crash-faulty replica (of an old synchronous group) might not transfer its latest state to a correct replica in the new synchronous group. This ``data loss'' fault is dangerous, as it may violate consistency when the system is in anarchy. However, such a fault can be detected using digital signatures from the commit log of some correct replicas (from an old synchronous group), provided that these correct replicas can  communicate synchronously with  correct replicas from the new synchronous group.
	In a sense, with \PCBFT FD, a critical non-crash machine fault must occur for the first time \emph{together} with sufficiently many crash or partitioned \processes (i.e., in anarchy) to violate consistency. 
	
	In the following, we explain the core of \PCBFT for the common case (Section~\ref{sec:CC}),  view-change (Section~\ref{sec:VC}) and fault detection (Section~\ref{sec:FD}) components. 
	We discuss \PCBFT optimizations in Section~\ref{sec:optimizations} and give 
	\PCBFT correctness arguments in Section~\ref{sec:correctness}. An example of \PCBFT execution is given in Appendix~\ref{apd:exp}. The complete pseudocode and correctness proof are included in Appendix~\ref{apd:pseudocode} and~\ref{apd:proof}. 
		
				\begin{figure}
		\centering
		\subfloat[$t\geq 2$]{\includegraphics[scale=0.80]{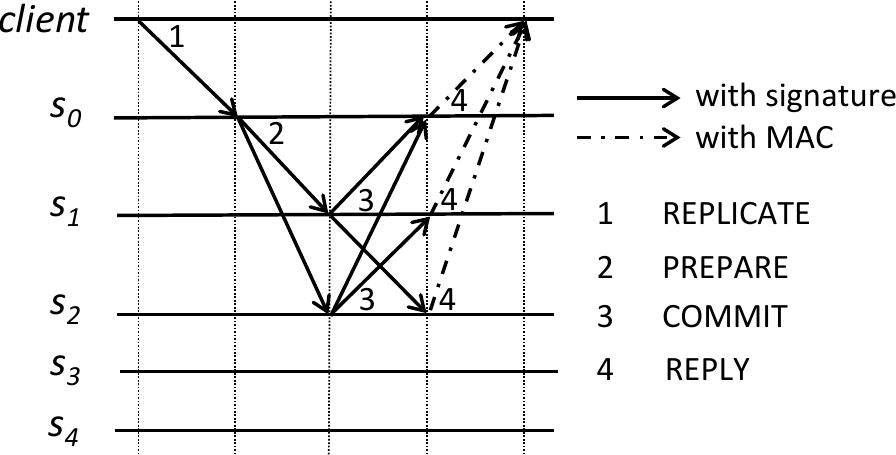}\label{fig:Atlas_f2}}
		\subfloat[$t=1$]{\includegraphics[scale=0.80]{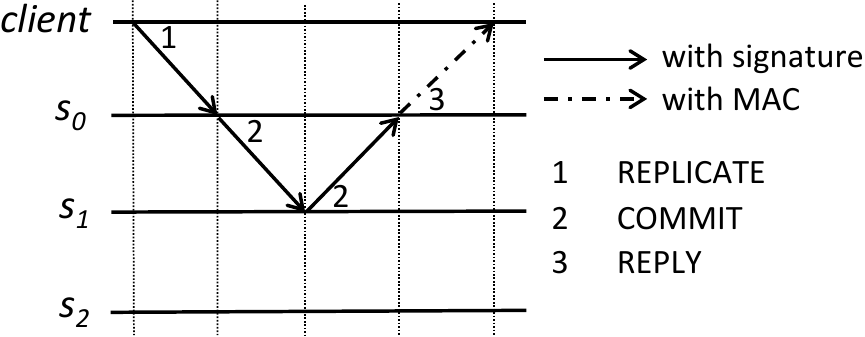}\label{fig:Atlas_f1}}
		
		\caption{\PCBFT\ common-case message patterns (a) for the general case when $t\geq 2$ and (b) for the special case of $t=1$. The synchronous groups are $(s_0, s_1, s_2)$ and $(s_0, s_1)$, respectively.}
		\label{fig:Atlas}
	\end{figure}

	\subsection{Common case}
	\label{sec:CC}
	
	Figure~\ref{fig:Atlas} shows the common-case message patterns of \PCBFT for the general case ($t\ge 2$) and for the special case $t=1$. \PCBFT is specifically optimized for the case where $t=1$, as in this case, there are only two active replicas in each view and the protocol is very efficient. The special case $t=1$ is also highly relevant in practice (see e.g., Spanner \cite{Corbett:2012:SGG:2387880.2387905}). In the following, we first explain \PCBFT in the  general case, and then focus on the $t=1$ special case.

	\bigskip
	
		 \noindent\textbf{Notation.} We denote the digest of a message $m$ by $D(m)$, whereas $\langle m\rangle_{\sigma_p}$ denotes a message that contains both $D(m)$ signed by the private key of \process $p$ and $m$. For signature verification, we assume that all \processes have public keys of all other processes.
	
	\subsubsection{General case ($t\ge 2$)}

 The common-case message pattern of \PCBFT is shown in Figure~\ref{fig:Atlas_f2}. More specifically, upon receiving a signed request $req=\langle \msgtag{replicate}, op,$ $ts_c, c\rangle_{\sigma_c}$  from client $c$ (where $op$ is the client's operation and $ts_c$ is the client's timestamp), the primary (say $s_0$) (1) increments sequence number $sn$ and assigns $sn$ to $req$, (2) signs a message $prep=\langle \msgtag{prepare}, D(req), sn, i\rangle_{\sigma_{s_0}}$ and logs $\angular{req,prep}$ into its prepare log $PrepareLog_0[sn]$ (we say $s_0$ \emph{prepares} $req$), and (3) forwards $\angular{req,prep}$ to \emph{all other active replicas} (i.e, the $t$ followers).
 
	Each follower $s_j$ ($1\leq j\leq t$) verifies the primary's and client's signatures, checks  whether its local sequence number equals $sn-1$, and logs $\angular{req,prep}$ into its prepare log $PrepareLog_j[sn]$. Then, $s_j$ updates its local sequence number to $sn$, \emph{signs} the digest of the request $req$, the sequence number $sn$ and the view number $i$, and sends $\langle \msgtag{commit}, D(req), sn, i\rangle_{\sigma_{s_j}}$ to all active replicas. 
	
	Upon receiving $t$ signed \msgtag{commit} messages --- one from each follower --- such that a matching entry is in the prepare log, an active replica $s_k$ ($0\leq k\leq t$) logs $prep$ and the $t$ signed \msgtag{commit} messages into its commit log $CommitLog_{s_k}[sn]$.  We say $s_k$ \emph{commits} $req$ when this occurs. Finally, $s_k$ executes $req$ and sends the authenticated reply to the client (followers may only send the digest of the reply). The client commits the request when it receives matching \msgtag{reply} messages from all $t+1$ active replicas.

	 A client that times out without committing the requests broadcasts the request to all active replicas. Active replicas then forward such a request to the primary and trigger a \emph{retransmission timer}, within which a correct active replica expects the client's request to be committed.

	\subsubsection{Tolerating a single fault ($t=1$).} When $t=1$, the \PCBFT\ common case simplifies to involving only 2 messages between 2 active replicas  (see Figure~\ref{fig:Atlas_f1}). 
	
	Upon receiving a signed request $req=\langle \msgtag{replicate}, op,$ $ts_c, c\rangle_{\sigma_c}$  from client $c$, the primary ($s_0$) increments the sequence number $sn$, signs $sn$ along the digest of $req$ and view number $i$ in message $m_0=\langle \msgtag{commit}, D(req), sn,$ $i\rangle_{\sigma_{s_0}}$, stores $\angular{req,m_0}$ into its prepare log ($PrepareLog_{s_0}[sn] = \langle req, m_0\rangle$), and sends the message $\angular{req,m_0}$ to the follower $s_1$.
	
	On receiving $\angular{req,m_0}$, the follower $s_1$ verifies the client's and primary's signatures, and checks whether its local sequence number equals $sn-1$.
	If so, the follower updates its local sequence number to $sn$, executes the request producing reply $R(req)$, and signs message $m_1$; $m_1$ is similar to $m_0$, but also includes the client's timestamp and the digest of the reply: $m_{1}=\langle \msgtag{commit}, \langle D(req), sn, i, req.ts_c,$ $ D(R(req))\rangle_{\sigma_{s_1}}$. The follower then saves the tuple $\langle req,m_0,m_1\rangle$ to its commit log ($CommitLog_{s_1}[sn]=\langle req,m_0,m_1\rangle$) and sends $m_1$ to the primary. 
	
	The primary, on receiving a valid \msgtag{commit} message from the follower (with a matching entry in its prepare log), executes the request, compares the reply $R(req)$ with the follower's digest contained in $m_1$, and stores $\langle req,m_0,m_1\rangle$ in its commit log. Finally, it returns an authenticated reply containing $m_1$ to $c$, which commits the request if all digests and the follower's signature match. 
	
	\subsection{View change}
	\label{sec:VC}
	
	\noindent\textbf{Intuition.} The ordered requests in commit logs of correct replicas are the key to enforcing consistency (total order) in \PCBFT.
	To illustrate an \PCBFT\ view change,  
	consider synchronous groups $\vs_i$ and $\vs_{i+1}$ of views $i$ and $i+1$, respectively, each containing $t+1$ replicas. Note that proofs of requests committed in $\vs_i$ might have been logged by \emph{only one} correct replica in $\vs_i$. Nevertheless, the \PCBFT\ view change must ensure that (outside anarchy) these proofs are transferred to the new view $i+1$. To this end, we had to depart from traditional view change techniques \cite{Castro:2002:PBF, Kotla:2009:ZSB,Clement:2009:MBF:1558977.1558988} where the entire view-change is led by a single replica, usually the primary of the new view.
	Instead, in \PCBFT view change, \emph{every active replica in $\vs_{i+1}$ retrieves information about requests committed in preceding views}.
	Intuitively, with correct majority of correct and synchronous replicas, at least one correct and synchronous replica from $\vs_{i+1}$ will contact (at least one) correct and synchronous replica from $\vs_i$ and transfer the latest correct commit log to the new view $i+1$.
	
	In the following, we first describe how we choose active replicas for each view. Then, we explain how view changes are initiated, and, finally, how view changes are performed. 
	
	\subsubsection{Choosing active replicas} 
	
	To choose active replicas for view $i$, we may enumerate all sets containing $t+1$ replicas (i.e., $\binom{2t+1}{t+1}$ sets) which then alternate as synchronous groups  across views in a round-robin fashion.
	In addition, each synchronous group uniquely determines the primary. We assume that the mapping from view numbers to synchronous groups is known to all replicas (see e.g., Table~\ref{table:sgcomb}).
	
	The above simple scheme works well for small number of replicas (e.g., $t=1$ and $t=2$). For a large number of replicas, the combinatorial number of synchronous groups may be inefficient. To this end, \PCBFT can be modified to rotate only the leader, which may then resort to deterministic verifiable pseudorandom selection of the set of $f$ followers in each view. The exact details of such a scheme would, however, exceed the scope of this paper.
	
	\begin{table}
		\center
		\small
		\begin{tabular}{|c|c|c|c|c|}
			\cline{3-5}
			\multicolumn{1}{c}{} & \multicolumn{1}{c|}{} & \multicolumn{3}{c|}{Synchronous Groups}\\
			\multicolumn{1}{c}{} & \multicolumn{1}{c|}{} & \multicolumn{3}{c|}{($i\in\mathbb{N}_0$)}\\
			\cline{3-5}
			\multicolumn{1}{c}{} & \multicolumn{1}{c|}{} & $\sg_i$ & $\sg_{i+1}$ & $\sg_{i+2}$\\
			\hline
			\multirow{2}{*}{Active replicas} & Primary & $s_0$ & $s_0$ & $s_1$ \\
			\cline{2-5}
			& Follower &  $s_1$ & $s_2$ & $s_2$ \\
			\cline{1-5}
			\multicolumn{2}{|c|}{Passive replica} & $s_2$ & $s_1$ & $s_0$ \\
			\hline
		\end{tabular}
		\caption{Synchronous group combinations ($t=1$).}%
		\label{table:sgcomb}
	\end{table}

	\subsubsection{View-change initiation} 
	\label{sec:vcinit}
	
	If a synchronous group in view $i$ (denoted by $\vs_i$) does not make progress, \PCBFT\ performs a view change. Only an active replica of $\vs_i$ may initiate a view change. 	
	An active replica $s_j\in \vs_{i}$ initiates a view change if (i) $s_j$ receives a message from another active replica that does not conform to the protocol (e.g.,
	an invalid
	signature), (ii) the retransmission timer at $s_j$ expires, 
	(iii) $s_j$ does not complete a view change to view $i$ in a timely manner, or (iv) $s_j$ receives a valid $\msgtag{suspect}$ message for view $i$ from another replica in $\vs_{i}$. Upon a view-change initiation, $s_j$ stops participating in the current view and sends $\langle \msgtag{suspect}, i, s_j\rangle_{\sigma_{s_j}}$ to \emph{all} other replicas.

	\subsubsection{Performing the view change} 
	
\begin{figure}
		\centering
		\includegraphics[scale=0.80]{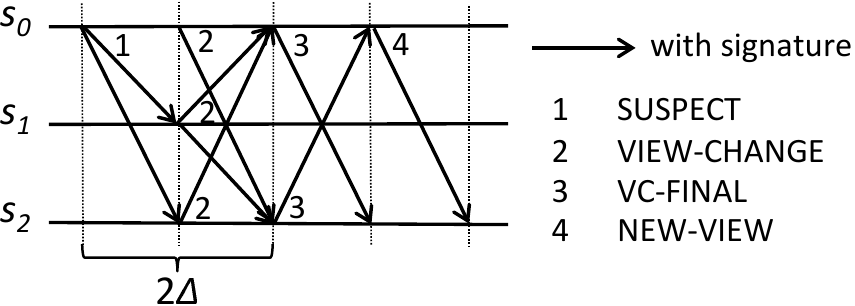}
		\caption{Illustration of \PCBFT\ view change: the synchronous group is changed from ($s_0$,$s_1$) to ($s_0$,$s_2$).}
		\label{fig:VRC}
	\end{figure}

	Upon receiving a \msgtag{suspect} message from an active replica in view $i$ (see the message pattern in Figure~\ref{fig:VRC}), replica $s_j$ stops processing messages of view $i$ and sends $m=\angular{\msgtag{view-change},i+1, s_j,$ $CommitLog_{s_j}}_{\sigma_{s_j}}$ to the $t+1$ active replicas of $sg_{i+1}$. A \msgtag{view-change} message contains the commit log $CommitLog_{s_j}$ of  $s_j$. Commit logs might be empty (e.g., if $s_j$ was passive). 
	
	Note that \PCBFT\ requires all active replicas in the new view to collect the most recent state and its proof (i.e., \msgtag{view-change} messages), rather than only the new primary. Otherwise, a faulty new primary could, even outside anarchy, purposely omit \msgtag{view-change} messages that contain the most recent state. Active replica $s_j$ in view $i+1$ waits for at least $n-t$ \msgtag{view-change} messages from all, but also waits for $2\timeout$ time, trying to collect as many messages as possible.
	
	Upon completion of the above protocol, each active replica $s_j\in\vs_{i+1}$ inserts all \msgtag{view-change} messages it has received into set $VCSet^{i+1}_{s_j}$. Then $s_j$ sends $\angular{\msgtag{vc-final},i+1,s_j,VCSet^{i+1}_{s_j}}_{\sigma_{s_j}}$ to every active replica in view $i+1$. This serves to exchange the received \msgtag{view-change} messages among active replicas.
	
	Every active replica $s_j\in\vs_{i+1}$ must receive \msgtag{vc-final} messages from \emph{all} active replicas in $\vs_{i+1}$, after which $s_j$ extends the value $VCSet^{i+1}_{s_j}$ by combining $VCSet^{i+1}_*$ sets piggybacked in \msgtag{vc-final} messages. Then, for each sequence number $sn$, an active replica selects the commit log with the highest view number in all \msgtag{view-change} messages, to confirm the committed request at $sn$. 
	
	Afterwards, to prepare and commit the selected requests in view $i+1$, the new primary $ps_{i+1}$ sends $\angular{\msgtag{new-view},i+1,PrepareLog}_{\sigma_{ps_{i+1}}}$ to every active replica in $\vs_{i+1}$, where the array $PrepareLog$ contains the prepare logs generated in view $i+1$ for each selected request. Upon receiving a \msgtag{new-view} message, every active replica $s_j\in\vs_{i+1}$ processes the prepare logs in $PrepareLog$ as described in the common case (see Section~\ref{sec:CC}).
	
	Finally, every active replica $s_j\in\vs_{i+1}$ makes sure that all selected requests in $PrepareLog$ are committed in view $i+1$. When this condition is satisfied, \PCBFT can start processing new requests.
	
	\subsection{Fault detection}
	\label{sec:FD}
	
	\PCBFT does not guarantee consistency in anarchy. Hence, non-crash faults could violate \PCBFT consistency in the long run, if they persist long enough to eventually coincide with enough crash or network faults. To cope with long-lived faults, we propose (an otherwise optional)  \emph{Fault Detection (FD)} mechanism for \PCBFT.
	  
	Roughly speaking, FD guarantees the following property: \emph{if a \process $p$ suffers a non-crash fault outside anarchy in a way that would cause inconsistency in anarchy, then \PCBFT\ FD detects $p$ as faulty (outside anarchy).} In other words, any potentially fatal fault that occurs outside anarchy would be detected by \PCBFT FD. 
	
	Here, we sketch how FD works in the case $t=1$ (see Section~\ref{sec:vcfd_pcode} for details), focusing on detecting a specific non-crash fault  that may render \PCBFT inconsistent in anarchy --- a \emph{data loss} fault by which a non-crash-faulty replica \emph{loses some of its commit log} prior to a view change. Intuitively, data loss faults are dangerous as they cannot be prevented by the straightforward use of digital signatures.

	Our FD mechanism entails modifying the \PCBFT view change as follows:
	in addition to exchanging their commit logs, replicas also exchange their prepare logs. Notice that in the case $t=1$ only the primary maintains a prepare log (see Section~\ref{sec:CC}).
	In the new view, the primary prepares and the follower commits all requests contained in transferred commit and prepare logs.
	
	With the above modification, to violate consistency, a faulty primary (of preceding view $i$) would need to exhibit a data loss fault in both its commit log and its prepare log. However, such a data loss fault in the primary's prepare log would be detected, outside anarchy, because (i) the (correct) follower of view $i$ would reply in the view change and (ii) an entry in the primary's prepare log causally precedes the respective entry in the follower's commit log. By simply verifying the signatures in the follower's commit log, the fault of a primary is detected. Conversely, a data loss fault in the commit log of the follower of view $i$ is detected outside anarchy by verifying the signatures in the commit log of the primary of view $i$.
	
	\subsection{\PCBFT optimizations}
	\label{sec:optimizations}

Although the common-case and view-change protocols described above are sufficient to guarantee correctness, we applied several standard performance optimizations to \PCBFT. These include checkpointing and lazy replication \cite{Lazy} to passive replicas (to help shorten the state transfer during view change) as well as batching and pipelining (to improve the throughput).
	
	\subsubsection{Checkpointing} 
	
\begin{figure}[!ht]
		\centering
		\includegraphics[scale=0.90]{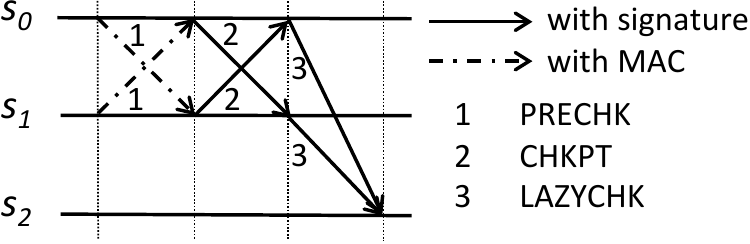}
		\caption{\PCBFT\ checkpointing message pattern : synchronous group is ($s_0$,$s_1$).}
		\label{fig:checkpointing}
	\end{figure}

	Upon active replica $s_j\in\vs_i$ commits and executes the  request with sequence number $sn=k\times CHK$ (refer to message pattern in Fig.~\ref{fig:checkpointing}) , $s_j$ sends $\angular{\msgtag{prechk},sn,$ $i,D(st^{sn}_{s_j}),s_j}_{\mu_{s_j,s_k}}$ to every active replica $s_k$, where $D(st^{sn}_{s_j})$ is the digest of the state after executing the request at $sn$. Upon receiving $t+1$ matching \msgtag{prechk} messages, each active replica $s_j$ generates the checkpoint proof message $m$ and sends it to every active replica ($m=\angular{\msgtag{chkpt}, sn, i,$ $ D(st^{sn}_{s_j}),s_j}_{\sigma_{s_j}}$). Upon receiving $t+1$ matching \msgtag{chkpt} messages, each active replica $s_j$ checkpoints the state and discards previous prepare logs and commit logs.
	
	Besides, each active replica propagates checkpoint proofs to all passive replicas by $\angular{\msgtag{lazychk},chkProof}$,
	where $chkProof$ contains $t+1$ \msgtag{chkpt} messages.

		\subsubsection{Lazy replication} 
		
		To speed up the state transfer in view change, the followers in synchronous group lazily propagate the commit log to every passive replica. With lazy replication, the new active replica, which might be the passive replica in preceding view, could only retrieve the missing state from others. 
	
\begin{figure}[!ht]
		\centering
		\subfloat[$t=1$]{\includegraphics[scale=0.80]{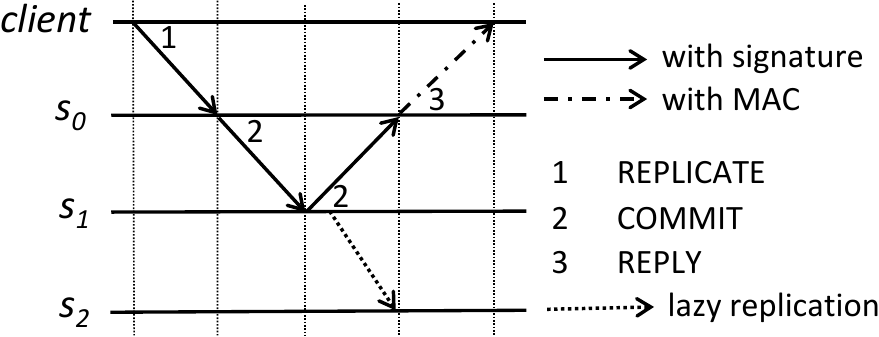}\label{fig:lazy_f1}}
		\subfloat[$t\ge 2$]{\includegraphics[scale=0.80]{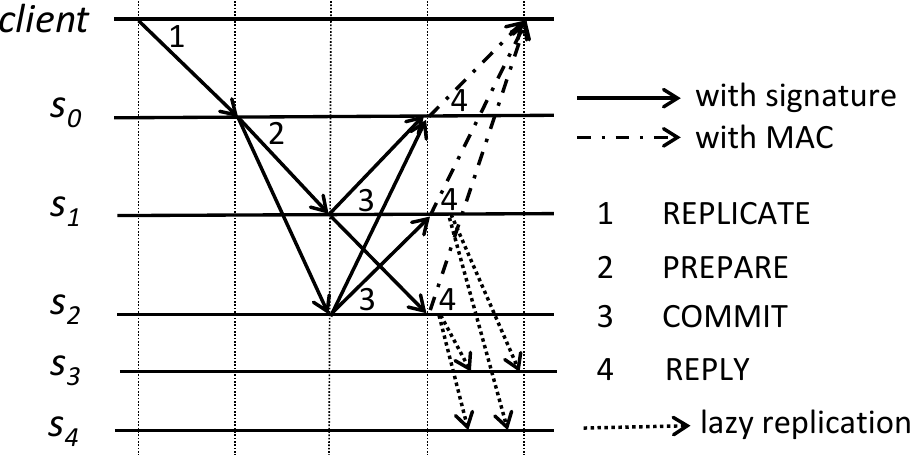}\label{fig:lazy_f2}}
		\caption{\PCBFT\ common-case message patterns with lazy replication for $t=1$ and $t\ge 2$ (here $t=2$). Synchronous group illustrated are ($s_0$,$s_1$) (when $t=1$) and ($s_0$,$s_1$,$s_2$) (when $t=2$), respectively.}
		\label{fig:lazy}
	\end{figure}

	More specifically, (refer to message pattern in Fig.~\ref{fig:lazy}) in case $t=1$, upon committing request $req$, the follower sends commit log of $req$ to the passive replica. In case $t\geq 2$, either each of $t$ followers sends commit log of $req$ to one passive replica, or each follower sends a fraction of $\frac{1}{t}$ commit logs to every passive replica. 
	Only in case the bandwidth between followers and passive replicas are saturated, the primary is involved in lazy replication. 
	Each passive replica commits and executes requests based on orders defined by commit logs. 
	
	Although non-crash faulty replicas can interfere with the lazy replication scheme, this would not impact the correctness of the protocol, but only slow down the view-change.
	
	\noindent\textbf{Batching and pipelining.} To improve the throughput of cryptographic operations, the primary batches several requests when preparing. The primary waits for $B$ requests, then signs the batched request and sends it to every follower. If primary receives less than $B$ requests within a time limit, the primary batches all requests it has received.
	
	\subsection{Correctness arguments}
	\label{sec:correctness}
	
	\noindent\textbf{Consistency (Total Order).} \PCBFT enforces the following invariant, which is key to total order.
	
	\begin{lemma}
		\label{lemma:main}
		Outside anarchy, if a benign client $c$ commits a request $req$ with sequence number $sn$ in view $i$, and a benign replica $s_k$ commits the request $req'$ with $sn$ in view $i'>i$, then $req=req'$.
	\end{lemma}
	
	A benign client $c$ commits request $req$ with sequence number $sn$ in view $i$ only after $c$  has received matching replies from $t+1$ active replicas in $\vs_i$. This implies that every benign replica in $\vs_i$  stores  $req$ into its commit log under sequence number $sn$. 
	In the following, we focus on the special case where: $i'=i+1$.
	This serves as the base step for the proof of Lemma~\ref{lemma:main} by induction across views which we give in~Section~\ref{apd:proof}.
	
	Recall that, in view $i'=i+1$,  
	all (benign) replicas from $\vs_{i+1}$ wait for $n-t=t+1$ \msgtag{view-change} messages containing commit logs transferred from  other replicas, as well as for the timer set to $2\timeout$ to expire. Then, replicas in $\vs_{i+1}$ exchange this information within \msgtag{vc-final} messages. Note that, outside anarchy, there exists at least one \emph{correct} and \emph{synchronous} replica in $\vs_{i+1}$, say $s_j$. Hence, a benign replica $s_k$ that commits $req'$ in view $i+1$ under sequence number $sn$ must have had received \msgtag{vc-final} from $s_j$. In turn, $s_j$ waited for $t+1$ \msgtag{view-change} messages (and timer $2\Delta$), so it received a \msgtag{view-change} message from some correct and synchronous replica $s_x\in \vs_i$ (such a replica exists in $\vs_i$ as at most $t$ replicas in $\vs_i$ are non-crash-faulty or partitioned). As $s_x$ stored $req$ under $sn$ in its commit log in view $i$, it forwards this information to $s_j$ in a \msgtag{view-change} message, and $s_j$ forwards this information to $s_k$ within a \msgtag{vc-final}. Hence $req=req'$ follows.
	
	\smallskip
	
	\noindent\textbf{Availability.}
	\PCBFT availability is guaranteed if the synchronous group contains only correct and synchronous replicas. With eventual synchrony, we can assume that, eventually, there will be no network faults. In addition, with all combinations of $t+1$ replicas rotating in the role of active replicas, \PCBFT\ guarantees that, eventually, view change in \PCBFT\ will complete with $t+1 $ \emph{correct} and \emph{synchronous} active replicas.

	\section{Performance Evaluation}
	\label{sec:evaluation}

	\begin{table*}[!htbp]
		\begin{tiny}
			\centering
			\makebox[\textwidth][c]{
			\begin{tabular}{|l|c|c|c|c|c|}
				\cline{2-6}
				\multicolumn{1}{c|}{} & {US West 1 (CA)} & {Europe (EU)} & {Tokyo (JP)} & {Sydney (AU)} & {Sao Paolo (BR)} \\
				\hline
				US East (VA)&
				88 /1097 /82190 /166390 &
				92 /1112 /85649 /169749 &
				179 /1226 /81177 /165277 &
				268 /1372 /95074 /179174 &
				146 /1214 /85434 /169534 \\
				\hline
				US West 1 (CA) &  &
				174 /1184 /1974 /15467 &
				120 /1133 /1180 /6210 &
				186 /1209 /6354 /51646 &
				207 /1252 /90980 /169080  \\
				\hline
				Europe (EU) & & &
				287 /1310 /1397 /4798 &
				342 /1375 /3154 /11052 &
				233 /1257 /1382 /9188 \\
				\hline
				Tokyo  (JP) & &  &  &
				137 /1149 /1414 /5228 &
				394 /2496 /11399 /94775 \\
				\hline
				Sydney (AU) & &  &   &  &
				392 /1496 /2134 /10983 \\
				\hline
			\end{tabular}}
		\end{tiny}
		\caption{Round-trip latency of TCP ping (\emph{hping3}) across Amazon EC2 datacenters, collected during three months. The latencies are given in milliseconds, in the format: average / 99.99\% / 99.999\% / maximum.}
\label{table:hping}
	\end{table*}
	
	In this section, we evaluate the performance of \PCBFT\ and compare it to that of Zyzzyva \cite{Kotla:2009:ZSB}, PBFT \cite{Castro:2002:PBF} and a WAN-optimized version of Paxos \cite{Lamport:1998:PP:279227.279229}, using the
	Amazon
	EC2 worldwide cloud platform. We chose geo-replicated, WAN settings as we believe that these are a better fit for protocols that tolerate Byzantine faults, including XFT and BFT. Indeed, in WAN settings \emph{(i)} there is no single point of failure such as a switch interconnecting machines, \emph{(ii)} there are no correlated failures due to, e.g., a power-outage, a storm, or other natural disasters, and  \emph{(iii)} it is difficult for the adversary to flood the network, correlating network and non-crash faults (the last point is relevant for XFT).
	
	In the remainder of this section, we first present the experimental setup (Section~\ref{sec:settings}), and then evaluate the performance (throughput, latency and CPU cost) in the fault-free scenario (Section~\ref{sec:latency} and Section~\ref{sec:cpu}) as well as under 
	faults (Section~\ref{sec:faults}). Finally, we perform a performance comparison using a real application,  the ZooKeeper coordination service \cite{Zookeeper} (Section~\ref{sec:ZK}), by comparing native ZooKeeper to ZooKeeper variants that use the four replication protocols mentioned above.

	\subsection{Experimental setup}
	\label{sec:settings}
	
	\subsubsection{Synchrony and \PCBFT}
	\label{sec:synchrony}
	
	In a practical deployment of \PCBFT, a critical parameter is the value of timeout $\Delta$, i.e., the upper bound on the communication delay between any two \emph{correct} machines. If the round-trip time (RTT) between two correct machines takes more than $2\Delta$, we declare a network fault (see Section~\ref{sec:xpaxos_model}). Notably, $\Delta$ is vital to the \PCBFT view-change (Section~\ref{sec:VC}).
	
	To understand the value of $\Delta$ in our geo-replicated context, we ran a 3-month experiment during which we continuously measured the round-trip latency across six Amazon EC2 datacenters worldwide using TCP ping (hping3). We used the least expensive EC2 micro instances, which  arguably have the highest probability of experiencing variable latency due to virtualization. Each instance was pinging all other instances every 100 ms. The results of this experiment are summarized in  Table~\ref{table:hping}. While we detected network faults lasting up to 3 min, our experiment showed that the round-trip  latency between \emph{any} two datacenters was less than 2.5 sec 99.99\% of the time. Therefore, we adopted the value of $\Delta = 2.5/2 = 1.25$ sec. 
	
	\subsubsection{Protocols under test}
	
\begin{figure}[htbp]
		\begin{center}
			\subfloat[PBFT]{\label{fig:pbft}\includegraphics[scale=0.60]{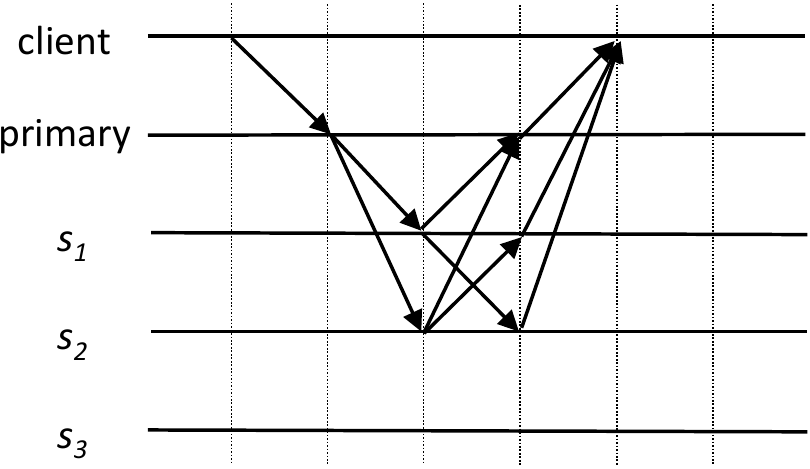}}
			\subfloat[Zyzzyva]{\label{fig:zyzzyva}\includegraphics[scale=0.60]{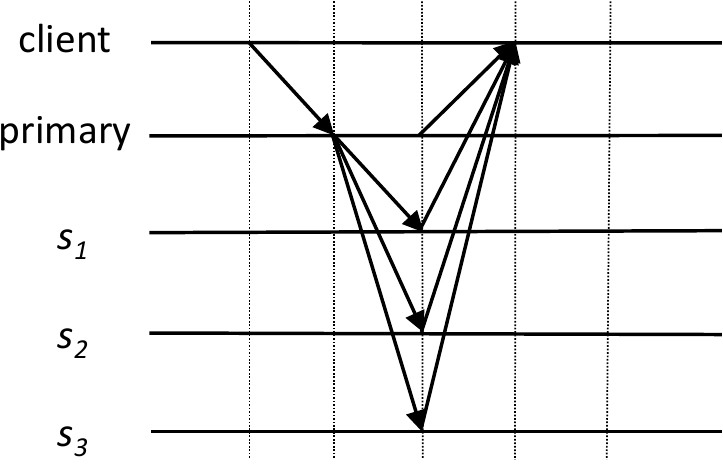}}
			\subfloat[Paxos]{\label{fig:paxos}\includegraphics[scale=0.65]{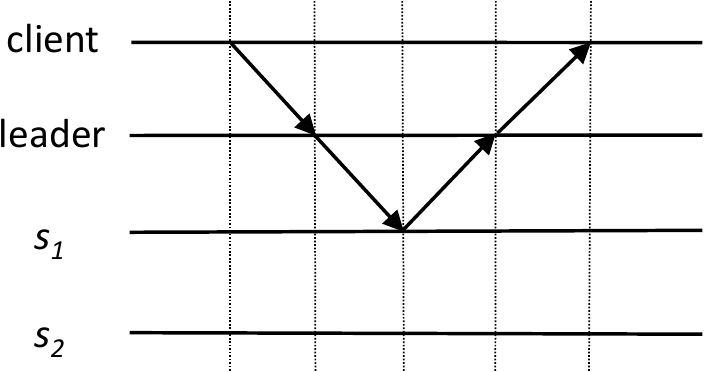}}
			\caption{Communication patterns of the three protocols under test ($t=1$).}
			\label{fig:crashpatterns}
		\end{center}
	\end{figure}

	We compare \PCBFT\ 
	 with three protocols whose common-case message patterns when $t=1$ are shown in Figure~\ref{fig:crashpatterns}. The first two are BFT protocols, namely (a speculative variant of) PBFT~\cite{Castro:2002:PBF} and Zyzzyva \cite{Kotla:2009:ZSB}, and require $3t+1$ replicas to tolerate $t$ faults. We chose PBFT because it is possible to derive a speculative variant of the protocol that relies on a 2-phase common-case commit protocol across only $2t+1$ replicas (Figure~\ref{fig:pbft}; see also \cite{Castro:2002:PBF}). In this PBFT variant, the remaining $t$ replicas
	are not involved in the common case, which is more efficient in a geo-replicated settings.
	We chose Zyzzyva because it is the fastest BFT protocol that involves all replicas in the common case (Figure~\ref{fig:zyzzyva}). 
	The third protocol we compare against is a very efficient WAN-optimized variant of crash-tolerant Paxos inspired by \cite{36971, MDCC, Corbett:2012:SGG:2387880.2387905}. 
	We have chosen the variant of Paxos that exhibits the fastest write pattern (Figure~\ref{fig:paxos}). This variant requires $2t+1$ replicas to tolerate $t$ faults, but involves $t+1$ replicas in the common case, i.e., just like \PCBFT.
	
	To provide a fair comparison, all protocols rely on the same Java code base and use batching, with the batch size set to 20. We  rely on HMAC-SHA1 to compute MACs and RSA1024 to sign and verify signatures computed using the Crypto++ \cite{Cryptopp} library that we interface with the various protocols using JNI.
	
	\subsubsection{Experimental testbed and benchmarks}
	\begin{sloppypar}
		
		We run the experiments on the Amazon EC2 platform which comprises widely distributed datacenters, interconnected by the Internet. Communications between datacenters have a low bandwidth and a high latency. We run the experiments on mid-range virtual machines that contain 8 vCPUs, 15 GB of memory, 2 x 80 GB SSD storage, and run Ubuntu Server 14.04 LTS (PV) with the Linux 3.13.0-24-generic x86\_64 kernel.
		
		In the case $t=1$, Table~\ref{table:replica_combination_worldwidecloud} gives the deployment of the different replicas at different datacenters, for each protocol  analyzed.
		Clients are always located in the same datacenter as the (initial) primary to better emulate what is done in modern geo-replicated systems where clients are served by the closest datacenter~\cite{Sovran:2011:TSG:2043556.2043592,
			Corbett:2012:SGG:2387880.2387905}.\footnote{In practice, modern geo-replicated system, like Spanner~\cite{Corbett:2012:SGG:2387880.2387905}, use hundreds of CFT SMR instances across different partitions to accommodate geo-distributed clients.}
		
		To stress the protocols, we run a microbenchmark that is similar to the one used in~\cite{Castro:2002:PBF,Kotla:2009:ZSB}. In this microbenchmark, each server replicates a {\it null} service (this means that there is no execution of requests). Moreover, clients issue requests in \emph{closed-loop}: a client waits for a reply to its current request before issuing a new request. The benchmark allows both the request size and the reply size to be varied. For space limitations, we only report results for two request sizes (1kB, 4kB) and one reply size (0kB). We refer to these microbenchmarks as 1/0 and 4/0 benchmarks, respectively.
		
		\subsection{Fault-free performance}
		\label{sec:latency}
		
\begin{figure}[!htbp]
		\centering
		\subfloat[1/0 benchmark, $t=1$]{\label{fig:10_nolgr}\includegraphics[scale=0.4]{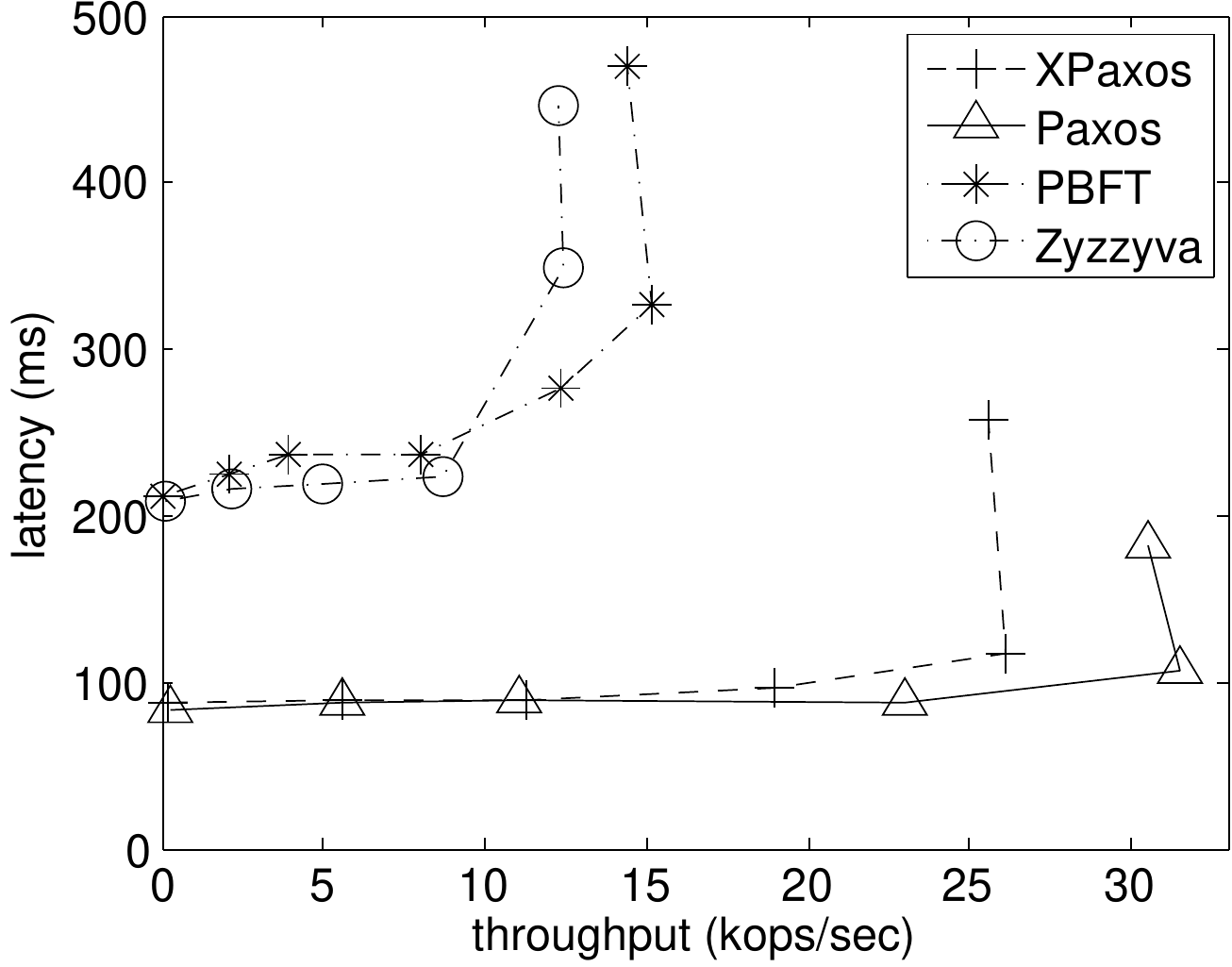}}
		\subfloat[4/0 benchmark, $t=1$]{\label{fig:40_nolgr}\includegraphics[scale=0.4]{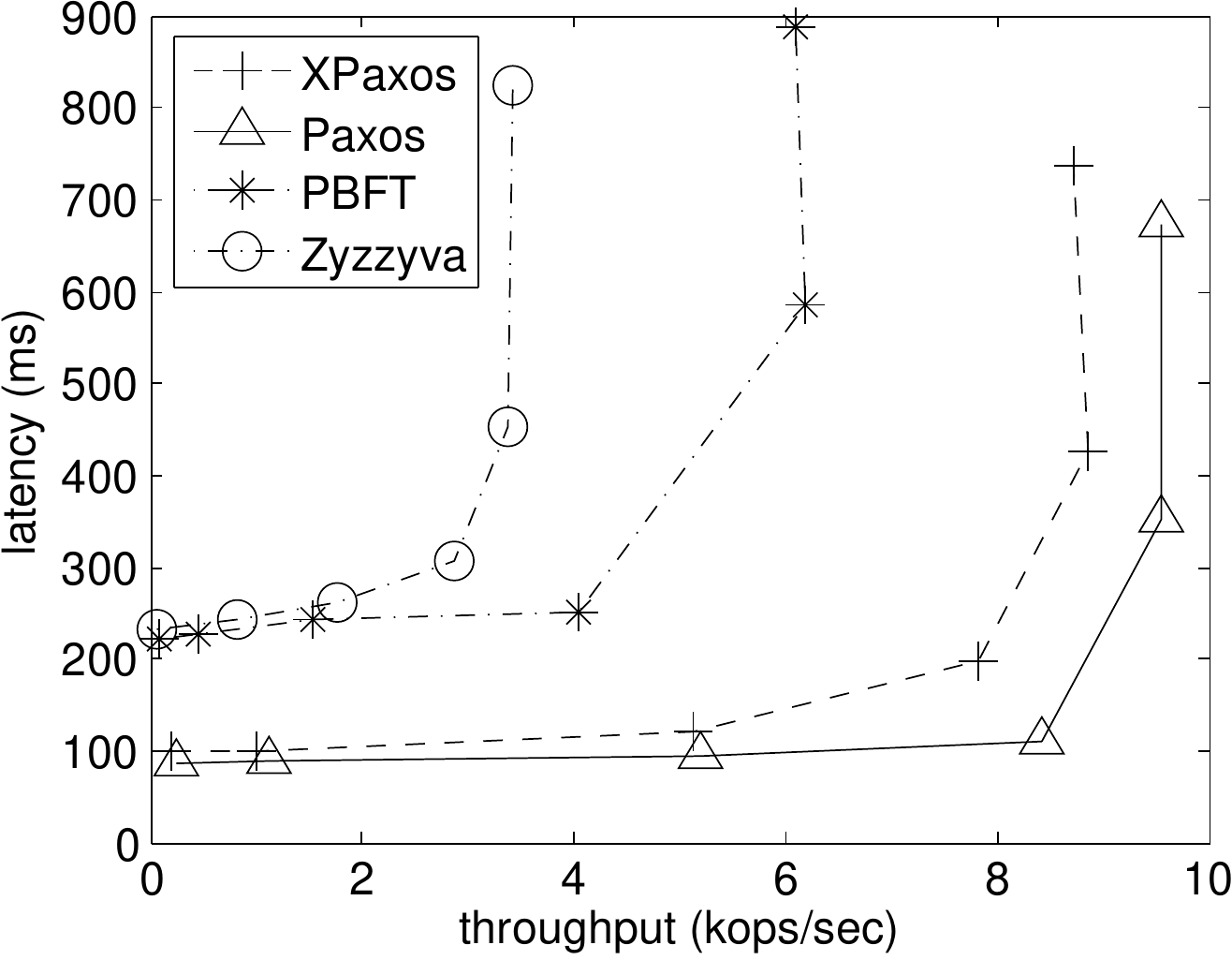}}
		\subfloat[1/0 benchmark, $t=2$]{		\label{fig:f2_10}
			\includegraphics[scale=0.4]{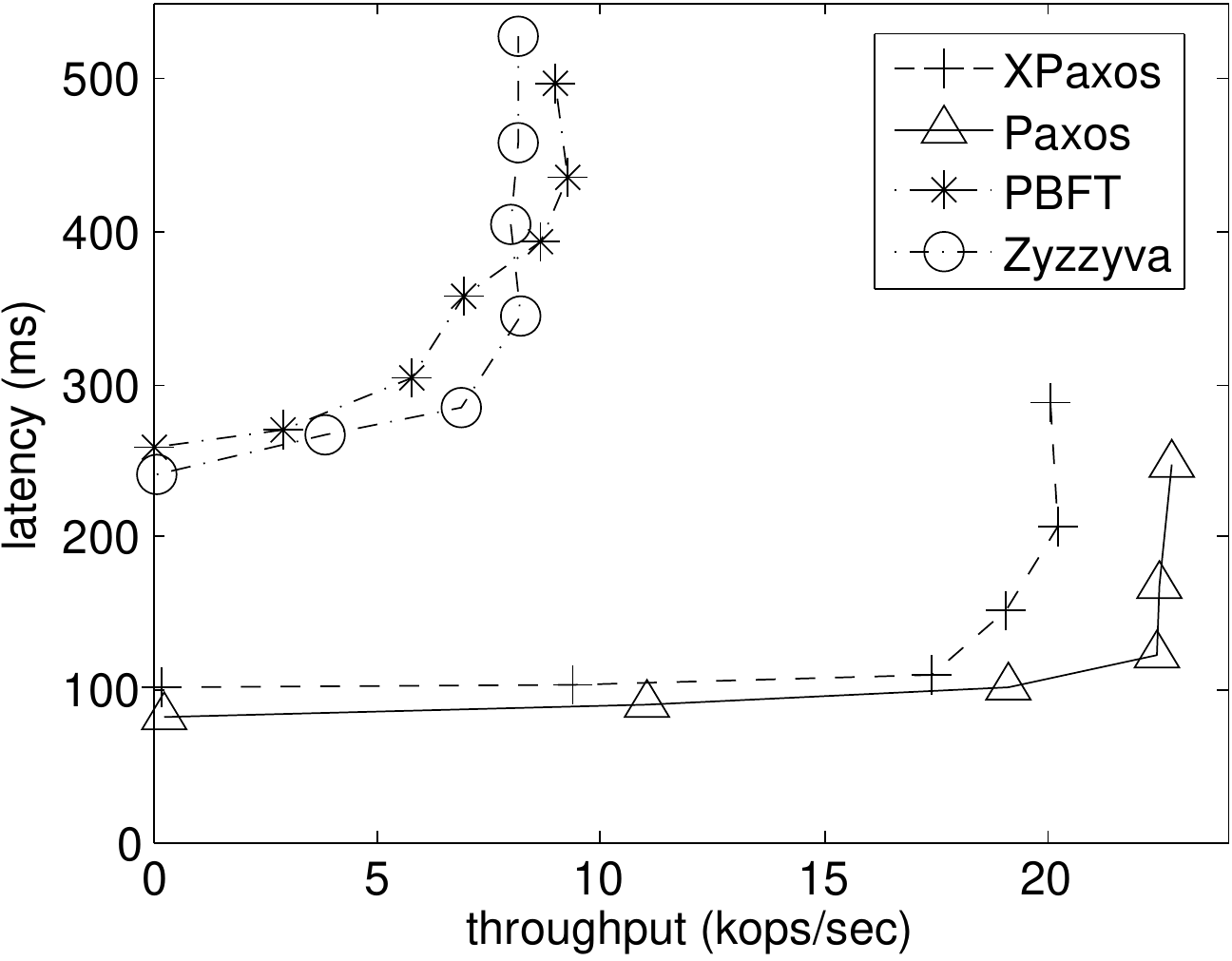}}
		\caption{Fault-free performance}
\label{fig:latency_vs_throughput_ec2_1}
	\end{figure}

		We first compare the performance of protocols when $t=1$  in replica configurations as shown in Table~\ref{table:replica_combination_worldwidecloud}, using the 1/0 and 4/0 microbenchmarks. The results are shown in Figures~\ref{fig:10_nolgr} and~\ref{fig:40_nolgr}.   In each graph, the X-axis shows the throughput (in kops/sec), and Y-axis the latency (in ms).
		
				\begin{table}[htbp]
			\begin{center}
				\begin{scriptsize}
					\begin{tabular}{|c|c|c|c||c|}
						\hline
						{\bf PBFT} & {\bf Zyzzyva} & {\bf Paxos} & {\bf \PCBFT} &{\bf EC2 Region}\\
						\hline
						Primary & Primary & Primary & Primary &  US West (CA)\\
						\hline
						\multirow{2}{*}{Active}   &\multirow{3}{*}{Active} & Active & Follower & US East (VA) \\
						\cline{3-3}\cline{4-5}
						&  & \cellcolor{lightgray}Passive & \cellcolor{lightgray}Passive&  Tokyo (JP) \\
						\cline{1-1}\cline{3-5}
						\cellcolor{lightgray}Passive & & - & - &   Europe (EU)\\
						\hline
					\end{tabular}
\caption{Configurations of replicas. Shaded replicas are not used in the common case.
					}
\label{table:replica_combination_worldwidecloud}
				\end{scriptsize}
			\end{center}
		\end{table}
				
		As we can see, in both benchmarks, \PCBFT\ achieves a significantly better performance than PBFT and Zyzzyva. This is because, in a worldwide cloud environment, the network is the bottleneck and the message patterns of BFT protocols, namely PBFT and Zyzzyva, tend to be expensive. Compared with PBFT, the simpler message pattern of \PCBFT allows better throughput.  Compared with Zyzzyva, \PCBFT puts less stress on the leader and replicates requests in the common case across 3 times fewer replicas than Zyzzyva (i.e., across $t$ followers vs. across all other $3t$ replicas). Moreover, the performance of \PCBFT  is very close to that of Paxos.  Both Paxos and \PCBFT\ implement a round-trip across two replicas when $t=1$, which renders them very efficient.

		Next, to assess the fault scalability of \PCBFT, we ran the 1/0 micro-benchmark in configurations that tolerate two faults ($t=2$). 
		We use the following EC2 datacenters for this experiment: CA (California), OR (Oregon), VA (Virginia), JP (Tokyo), EU (Ireland), AU (Sydney) and SG (Singapore).
		We place Paxos and \PCBFT\ active replicas in the first $t+1$ datacenters, and their passive replicas in the next $t$ datacenters. PBFT uses the first $2t+1$ datacenters for active replicas and the last $t$ for passive replicas. Finally, Zyzzyva uses all replicas as active replicas.  
		
		We observe that \PCBFT again clearly outperforms PBFT and Zyzzyva and achieves a performance very close to that of Paxos. 
		Moreover, unlike PBFT and Zyzzyva, Paxos and \PCBFT\ only suffer a moderate performance decrease with respect to the $t=1$ case.
		
	\end{sloppypar}

\subsection{CPU cost}
\label{sec:cpu}
	
	\begin{figure}
		\centering
		\includegraphics[scale=0.35]{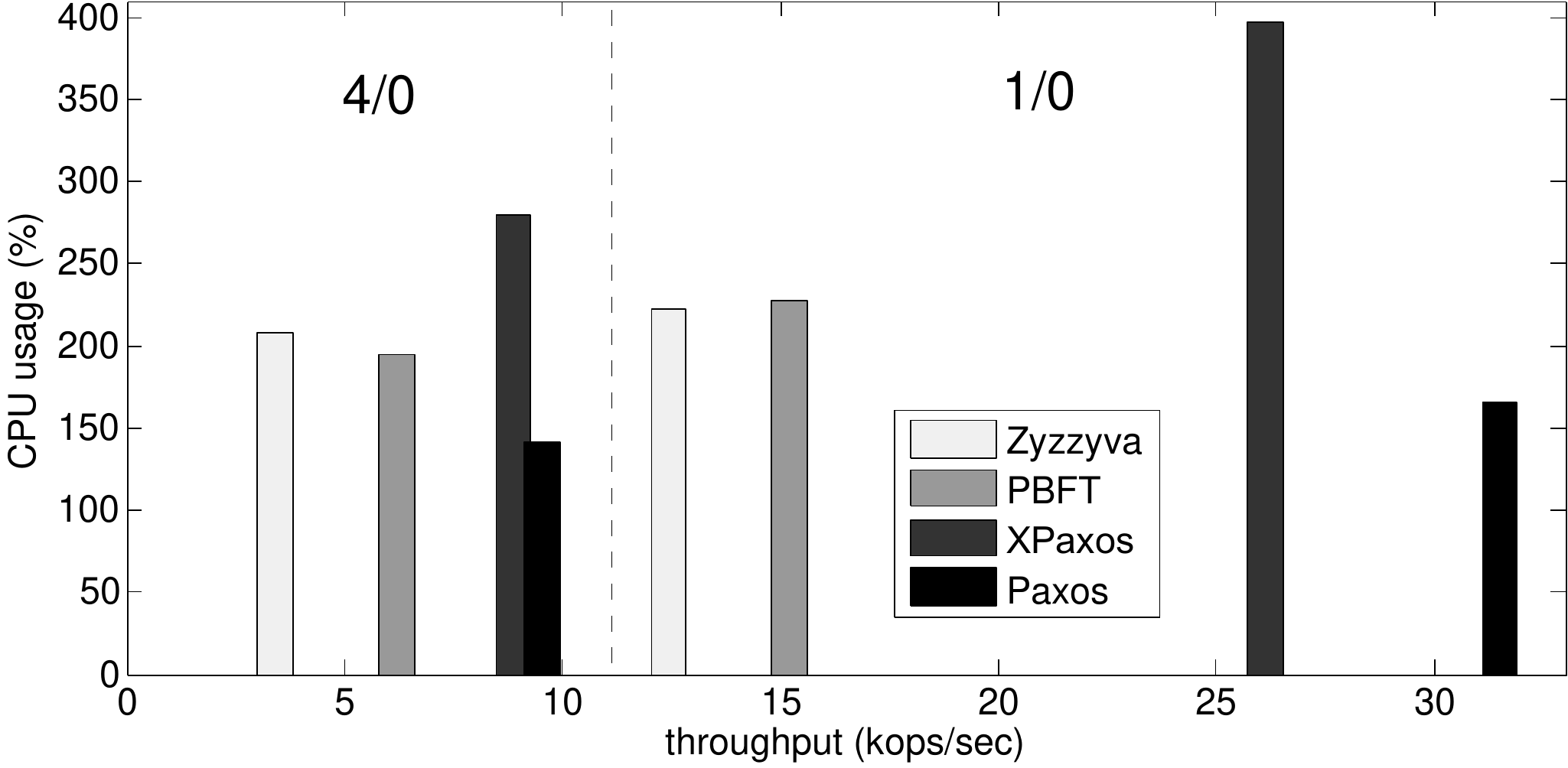}
		\caption{CPU usage when running the 1/0 and 4/0 micro-benchmarks.}
		\label{fig:40_cpu}
	\end{figure}

	To assess the cost of using signatures in \PCBFT, we extracted the CPU usage during the experiments presented in Section~\ref{sec:latency} with
	1/0 and 4/0 micro-benchmarks when $t=1$. During experiments, we periodically sampled CPU usage at the most loaded node (the primary in every protocol) with the \emph{top} Linux monitoring tool. The results are depicted in Figure~\ref{fig:40_cpu} for both the 1/0 and 4/0 micro-benchmarks. The X-axis represents the peak throughput (in kops/s), whereas the Y-axis represents the CPU usage (in \%).  Not surprisingly, we observe that the CPU usage of all protocols is higher with the 1/0 benchmark than with the 4/0 benchmark. This comes from the fact that in the former case, there are more messages to handle per time unit. We also observe that the CPU usage of \PCBFT\ 
	is higher than that of other protocols, due to the use of digital signatures. Nevertheless, this cost remains very reasonable: never more than half of the eight cores available on the experimental machines were used. 
	Note that this cost could probably be significantly reduced by using GPUs, as recently proposed on the EC2 platform. Moreover, compared to BFT protocols (PBFT and Zyzzyva), while CPU usage of \PCBFT\ is higher,
	\PCBFT\ also sustains a significantly higher throughput.
	
	\subsection{Performance under faults}
	\label{sec:faults}
	
\begin{figure}[htbp]
		\begin{center}
			\includegraphics[scale=0.35]{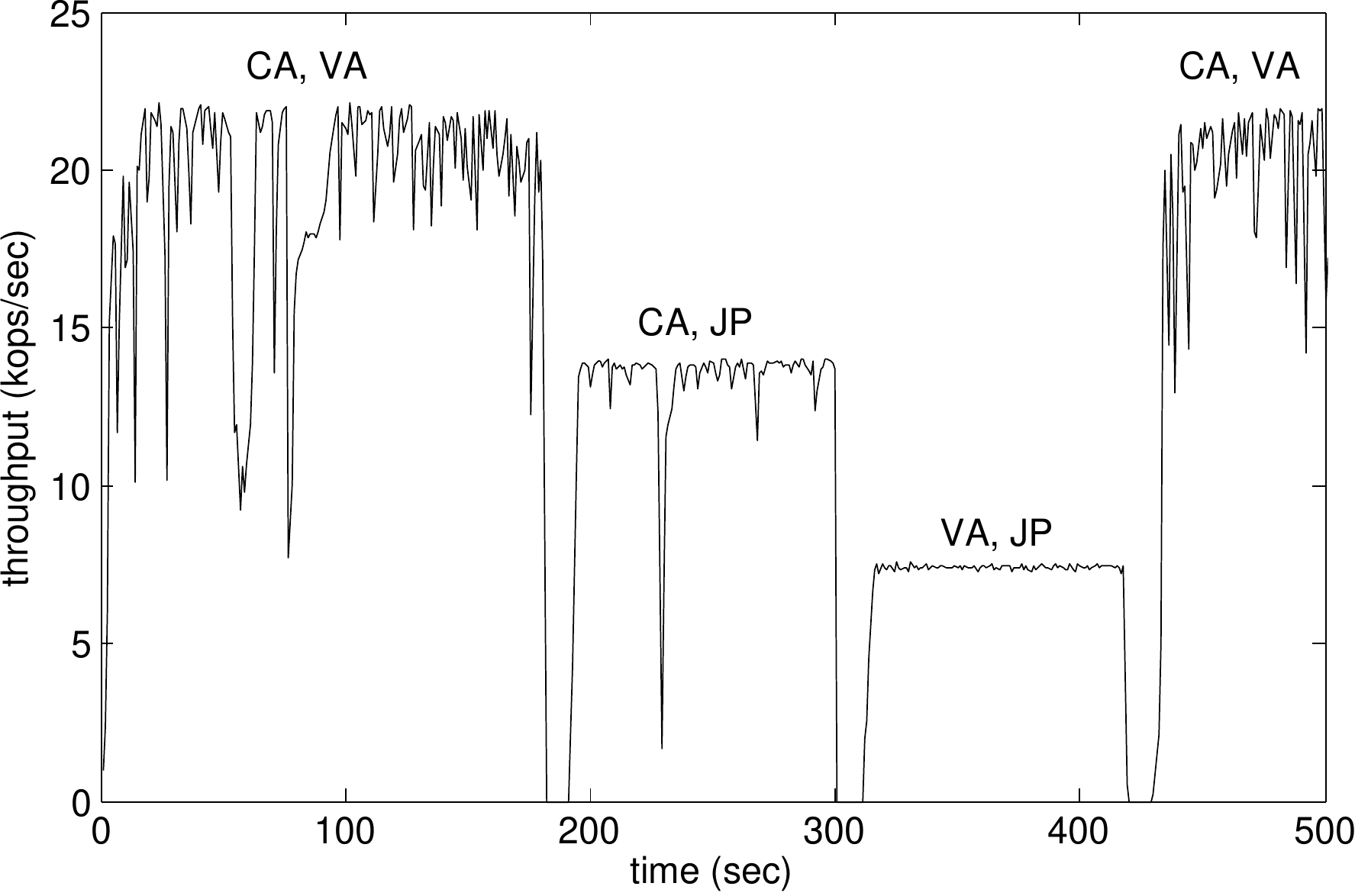} \end{center}
		\caption{\PCBFT under faults.}
		\label{fig:fault} \end{figure}

	In this section, we analyze the behavior of \PCBFT under faults. We run the 1/0 micro-benchmark on three replicas (CA, VA, JP) to tolerate one fault (see also  Table~\ref{table:replica_combination_worldwidecloud}). The experiment starts with CA and VA as active replicas, and with 2500 clients in CA. At time 180 sec, we crash the follower, VA. At time 300 sec, we crash the CA replica. At time 420 sec, we crash the third replica, JP. Each replica recovers 20 sec after having crashed. Moreover, the timeout $2\Delta$ (used during state transfer in view change, Section~\ref{sec:VC}) is set to 2.5 sec (see Section~\ref{sec:synchrony}). We show the throughput of \PCBFT in  function of time in Figure~\ref{fig:fault}, which also indicates the active replicas for each view. We observe that after each crash, the system performs a view change that lasts less than 10 sec, which is very reasonable in a geo-distributed setting. This fast execution of the view-change subprotocol is a consequence of lazy replication in  \PCBFT that keeps passive replicas updated. We also observe that the throughput of \PCBFT changes with the views. This is because the latencies between the primary and the follower and between the primary and clients vary  from view to view.
	
	\subsection{Macro-benchmark: ZooKeeper}
	\label{sec:ZK}
	
	To assess the impact of our work on real-life applications, we measured the performance achieved when replicating the ZooKeeper coordination service~\cite{Zookeeper} using all protocols considered in this study: Zyzzyva, PBFT, Paxos and \PCBFT. We also compare with the native ZooKeeper performance, when the system is replicated using the built-in \emph{Zab} protocol~\cite{Zab}. This protocol is crash-resilient and requires $2t+1$ replicas to tolerate $t$ faults.
	
	We used the ZooKeeper 3.4.6 codebase. The integration of the various protocols inside ZooKeeper was carried out by replacing the Zab protocol.
	For fair comparison to native ZooKeeper, we made a minor modification to native ZooKeeper to force it to use (and keep) a given node as primary. 	To focus the comparison on the performance of replication protocols, and avoid hitting other system bottlenecks (such as storage I/O that is not very efficient in virtualized cloud environments), we store ZooKeeper data and log directories on a volatile \emph{tmpfs} file system. 
	The  configuration tested tolerates one fault ($t=1$).
	ZooKeeper clients were 	located in the same region as the primary (CA).
	Each client invokes 1 kB write operations in a closed loop.
	
\begin{figure}[!htbp]
		\centering
		\includegraphics[scale=0.4]{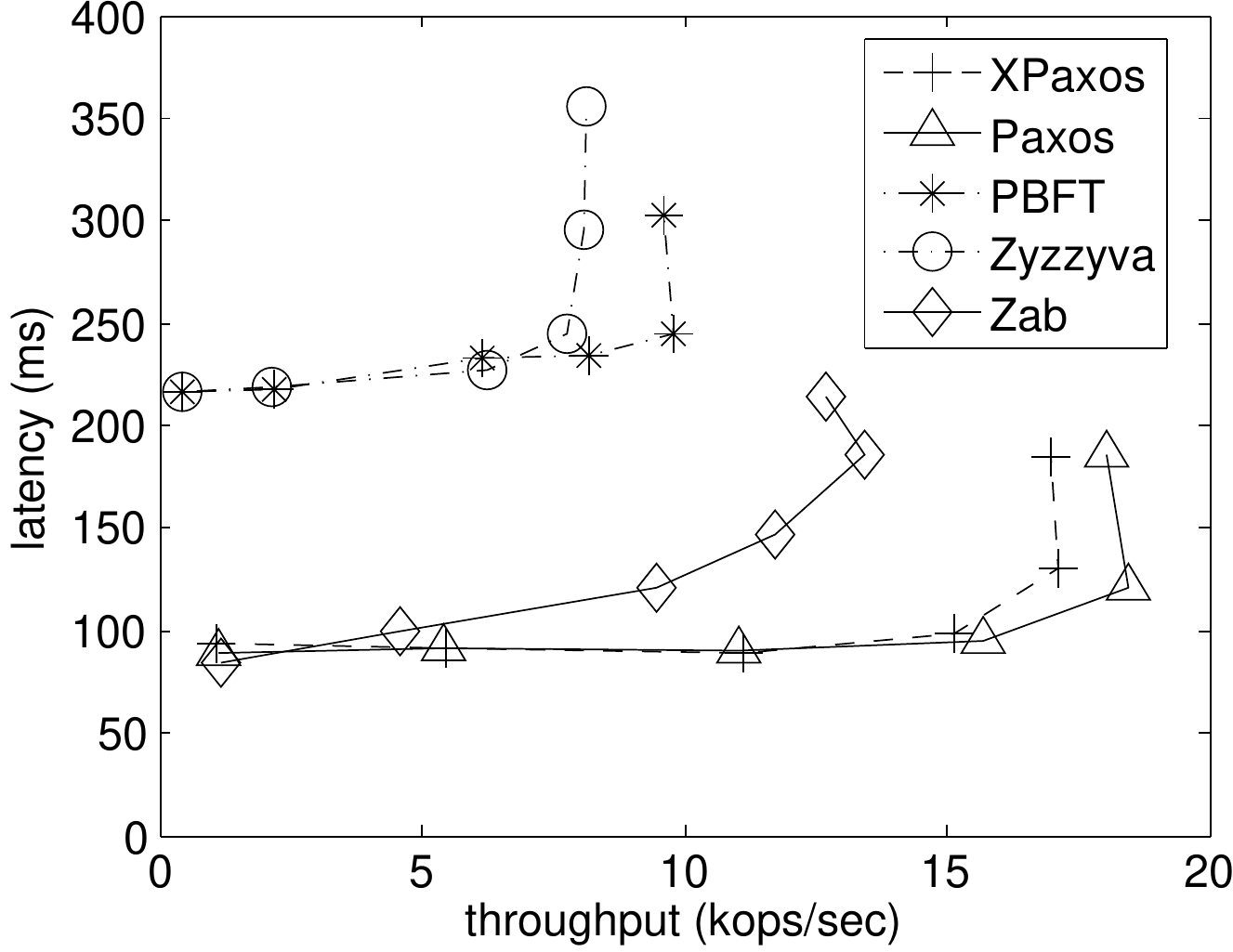}
		\caption{Latency vs. throughput for the ZooKeeper application ($t = 1$).}
		\label{fig:zk-thr_lat}
	\end{figure}

	Figure~\ref{fig:zk-thr_lat} depicts the results. The X-axis represents the throughput in kops/sec. The Y-axis represents the latency in ms. In this macro-benchmark, we find that Paxos and \PCBFT\ clearly outperform BFT protocols and that \PCBFT\ achieves a performance close to that of Paxos. More surprisingly, we can see that \PCBFT\ is more efficient than the built-in Zab protocol, although the latter only tolerates crash faults. For both protocols, the bottleneck in the WAN setting is the bandwidth at the leader, but the leader in Zab sends requests to all other $2t$ replicas whereas the \PCBFT leader sends requests only to $t$ followers, which yields a higher peak throughput for \PCBFT.
	
		\section{Reliability Analysis}
	\label{sec:nines}

In this section, we illustrate the reliability guarantees of \PCBFT by analytically comparing them with those of the state-of-the-art asynchronous CFT and BFT protocols. For simplicity of the analysis,  
		we consider the fault states of the \processes to be independent and identically distributed random variables. 
			
	We denote the probability that a \replica is correct (resp., crash faulty) by $\pcorrect$ (resp., $\pcrash$).
	The probability that a \replica is benign is $\pbenign=\pcorrect+\pcrash$. Hence, a  \replica is non-crash faulty with probability~$\pnoncrash=1-\pbenign$. 
	
	Besides, we assume there is a probability $\psynchrony$ that a \replica is synchronous, where $\psynchrony$ is a function of $\timeout$, the network, and the system environment. Therefore, the probability that a \replica is partitioned equals $1-\psynchrony$.
	
	Based on the assumption that network faults and machine faults occur independently, it is straightforward to reason for a given \process, $\pbenign$ and $\pcorrect$ are independent from $\psynchrony$. Hence, the probability that a \process is available (i.e., correct and synchronous) is $\pnormal=\pcorrect\times\psynchrony$.
	
Aligned with the industry practice, we measure reliability guarantees and coverage of fault scenarios using \emph{nines of reliability}. Specifically, we distinguish \emph{nines of consistency} and \emph{nines of availability} and use these measures to compare different fault models. We introduce a function $\ninesOf(p)$ that turns a probability~$p$ into the corresponding number of ``nines'', by letting $\ninesOf(p)=\lfloor -\log_{10}(1-p)\rfloor$. For example, $\ninesOf(0.999) = 3$. For brevity, $\ninesbenign$ stands for $\ninesOf(\pbenign)$, and so on, for other probabilities of interest. Beyond the analysis and examples that follow, Appendix~\ref{app:reliability} contains additional examples of practical values of nines of reliability achieved by XFT, CFT and BFT protocols.
	
	\subsection{Consistency}

	We start with the number of \emph{nines of consistency} for an asynchronous CFT protocol, denoted by~$\ninesofC(CFT)=\ninesOf(P[\text{CFT is consistent}])$. As  $P[\text{CFT is consistent}]=\pbenign^n$, a straightforward calculation yields:
	\[\ninesofC(CFT)=\Big\lfloor -\log_{10}(1-\pbenign)-\log_{10}(\sum\limits_{i=0}^{n-1}\pbenign^i)\Big\rfloor,
	\]
	which gives $\ninesofC(CFT)\approx \ninesbenign-\lceil \log_{10}(n)\rceil$ for values of $\pbenign$ close to~1, when $\pbenign^i$ decreases slowly. As a rule of thumb, for small values of $n$, i.e., $n < 10$, we have  $\ninesofC(CFT) \approx \ninesbenign-1$.
	
	In other words, in typical configurations, where few faults are tolerated \cite{Corbett:2012:SGG:2387880.2387905}, a CFT system as a whole loses one nine of consistency from the likelihood that a single replica is benign.
	
	\subsubsection{\PCBFT vs.\ CFT} We now quantify the advantage of \PCBFT over asynchronous CFT. From Table.~\ref{table:comp}, if there is no non-crash fault, or there are no more than $t$ faults (machine faults or network faults), \PCBFT is consistent, i.e.,
	
	\[
	P[\text{\PCBFT is consistent}]=\pbenign^n+\sum\limits_{i=1}^{t=\lfloor \frac{n-1}{2}\rfloor}\binom{n}{i}\pnoncrash^i\times\]
	\[\sum\limits_{j=0}^{t-i}\binom{n-i}{j}
	\pcrash^j\times\pcorrect^{n-i-j}\times
	\sum\limits_{k=0}^{t-i-j}\binom{n-i-j}{k}\psynchrony^{n-i-j-k}\times
	(1-\psynchrony)^{k}.
	\]
	
	\newcommand{\psyncall}{\ensuremath{p_{sync\_all}}}
	
	To quantify the difference between \PCBFT and CFT more tangibly, we calculated $\ninesofC(\PCBFT)$ and  $\ninesofC(CFT)$ for all values of $\ninesbenign$, $\ninescorrect$ and $\ninessynchrony$ ($\ninesbenign\geq\ninescorrect$) between 1 and 20 in the special cases where  $t=1$ and $t=2$, which are  most relevant in practice. For $t=1$, we observed the following relation:
	
	\[
	\ninesofC(\PCBFT_{t=1})-\ninesofC(CFT_{t=1})=\]
	\[
	\begin{cases}
	\ninescorrect-1,& \ninesbenign>\ninessynchrony \text{ and } \\ & \ninessynchrony=\ninescorrect,\\
	min(\ninessynchrony,\ninescorrect), & \text{otherwise.}
	\end{cases}
	\]
	
		\[
	\ninesofC(\PCBFT_{t=2})-\ninesofC(CFT_{t=2})=\]
	\[
	\begin{cases}
	2\times\ninescorrect-2,& \ninesbenign>\ninessynchrony \text{ and } \\ & \ninessynchrony=\ninescorrect>1,\\
	2\times\ninescorrect,& \ninessynchrony>2\times\ninesbenign \text{ and } \\ & \ninesbenign=\ninescorrect,\\
	2\times min(\ninessynchrony,\ninescorrect)-1,&   \text{otherwise.}
	\end{cases}
	\]

	Hence, for $t=1$ we observe that the number of nines of consistency \PCBFT adds on top of CFT is proportional to the nines of probability for correct or synchronous \process. The added nines are not directly related to $\pbenign$, although $\pbenign\geq\pcorrect$ must hold. 
	
	\bigskip
	
	\noindent\emph{Example 1.} When $\pbenign=0.9999$ and $\pcorrect=\psynchrony=0.999$, we have $\pnoncrash=0.0001$ and $\pcrash=0.0009$. In this example,  $9\times\pnoncrash=\pcrash$, i.e., if a machine  suffers a faults 10 times, then one of these is a non-crash fault and the rest are crash faults. In this case,  $\ninesofC(CFT_{t=1})=\ninesbenign-1=3$, whereas $\ninesofC(\PCBFT_{t=1})-\ninesofC(CFT_{t=1})=
	\ninescorrect-1=2$, i.e., $\ninesofC(\PCBFT_{t=1})=5$. \PCBFT adds 2 nines of consistency on top of CFT and achieves 5 nines of consistency in total.
	
	\bigskip

	\noindent\emph{Example 2.} In a slightly different example, let $\pbenign=\psynchrony=0.9999$ and $\pcorrect=0.999$, i.e., the network behaves more reliably than in Example 1.  $\ninesofC(CFT_{t=1})=\ninesbenign-1=3$, whereas $\ninesofC(\PCBFT_{t=1})-\ninesofC(CFT_{t=1})=
	\pcorrect=3$, i.e., $\ninesofC(\PCBFT_{t=1})=6$. \PCBFT adds 3 nines of consistency on top of CFT and achieves 6 nines of consistency in total.
	
	\subsubsection{\PCBFT vs.\ BFT}
	
	Recall that (see Table~\ref{table:comp}) SMR in asynchronous BFT model is consistent whenever no more than one-third machines are non-crash faulty. Hence, 
	
	\[
	P[\text{BFT is consistent}]=
	\sum\limits_{i=0}^{t=\lfloor\frac{n-1}{3}\rfloor}\binom{n}{i}(1-\pbenign)^i\times\pbenign^{n-i}.
	\]
	
	\bigskip
	
	We first examine the conditions under which \PCBFT has stronger consistency guarantees than BFT. Fixing the value $t$ of tolerated faults, we observe that $P[\text{\PCBFT is consistent}]  > P[\text{BFT is consistent}]$ is equivalent to:
	
	\[
	\pbenign^{2t+1}+\sum\limits_{i=1}^{t}\binom{2t+1}{i}\pnoncrash^i\times\sum\limits_{j=0}^{t-i}\binom{2t+1-i}{j}\pcrash^j\times\]
	\[\pcorrect^{2t+1-i-j}\times
	\sum\limits_{k=0}^{t-i-j}\binom{2t+1-i-j}{k}\psynchrony^{2t+1-i-j-k}\times\]
	\[(1-\psynchrony)^{k}>
	\sum\limits_{i=0}^{t}\binom{3t+1}{i}\pbenign^{3t+1-i}(1-\pbenign)^i.
	\]
	
	In the special case when $t=1$, the above inequality simplifies to \[
	\pnormal>\pbenign^{1.5}.
	\]

	Hence, for $t=1$, \PCBFT has \emph{stronger consistency guarantees} than \emph{any} asynchronous BFT protocol whenever the probability that a machine is available is larger than 1.5 power of the probability that a machine is benign. This is despite the fact that BFT is more expensive than \PCBFT as $t=1$ implies $4$ replicas for BFT and only $3$ for \PCBFT.

	In terms of nines of consistency, again for $t=1$ and $t=2$, we calculated the difference in consistency between \PCBFT and BFT SMR, for all values of $\ninesbenign$,  $\ninescorrect$  and $\ninessynchrony$ ranging between 1 and 20, and observed the following relation: 
	
	\smallskip
	
	\[
	\ninesofC(BFT_{t=1})-\ninesofC(\PCBFT_{t=1})=\]
	\[
	\begin{cases}
	\ninesbenign-\ninescorrect+1,& \ninesbenign>\ninessynchrony \text{ and }\\ & \ninessynchrony=\ninescorrect,\\
	\ninesbenign-min(\ninescorrect,\ninessynchrony),& \text{otherwise.}
	\end{cases}
	\]
	
		\[
	\ninesofC(BFT_{t=2})-\ninesofC(\PCBFT_{t=2})=\]
	\[
	\begin{cases}
	2\times(\ninesbenign-\ninescorrect)+1,& \ninesbenign>\ninessynchrony \text{ and } \\ & \ninessynchrony=\ninescorrect,\\
	-1,& \ninessynchrony>2\times\ninesbenign \\ & \text{ and } \ninesbenign=\ninescorrect,\\
	2\times(\ninesbenign-min(\ninescorrect,\ninessynchrony)),& \text{otherwise.}
	\end{cases}
	\]
	
	Note that in cases where \PCBFT guarantees better consistency than BFT ($\pnormal>\pbenign^{1.5}$), it is only ``slightly'' better and does not materialize in additional nines.

	\smallskip
	\smallskip
	
	\noindent \emph{Example 1 (cont'd.).} Building upon our example,  $\pbenign=0.9999$ and $\psynchrony=\pcorrect=0.999$, we have $\ninesofC(BFT_{t=1})-\ninesofC(\PCBFT_{t=1})=
	\ninesbenign-\ninessynchrony+1=2$, i.e., $\ninesofC(\PCBFT_{t=1})=5$ and $\ninesofC(BFT_{t=1})=7$. BFT brings 2 nines of consistency on top of \PCBFT. 
	
	\smallskip
	\smallskip
	
	\noindent \emph{Example 2 (cont'd.).} When $\pbenign=\psynchrony=0.9999$ and $\pcorrect=0.999$, we have $\ninesofC(BFT_{t=1})-\ninesofC(\PCBFT_{t=1})=1$, i.e., $\ninesofC(\PCBFT_{t=1})=6$ and $\ninesofC(BFT_{t=1})=7$. \PCBFT has one nine of consistency less than BFT (albeit the only 7th).
	
	\subsection{Availability}
	
	Then, we quantify the stronger availability guarantees of \PCBFT over asynchronous CFT and BFT protocols.
We define the number of \emph{nines of availability} for protocol $X$, as~$\ninesofA(X)=\ninesOf(P[\text{X is available}])$. 
	
	Recalling that whenever  $\lfloor\frac{n-1}{2}\rfloor+1$ active replicas in synchronous group are available, \PCBFT can make progress despite passive replicas are benign or not, partitioned or not (see Table~\ref{table:comp}). Thus, we have $P[\text{\PCBFT is available}]=
	\sum\limits_{i=\lfloor\frac{n-1}{2}\rfloor+1}^{n}\binom{n}{i}\pnormal^{i}\times(1-\pnormal)^{n-i}$.
	
	\smallskip
	
	\subsubsection{\PCBFT vs.\ CFT}
	
	a CFT protocol (e.g., Paxos) is available whenever $n-\lfloor\frac{n-1}{2}\rfloor$ machines are correct and synchronous, plus other machines are benign (see Table~\ref{table:comp}). Hence, $P[\text{CFT is available}]=
	\sum\limits_{i=n-\lfloor\frac{n-1}{2}\rfloor}^{n}\binom{n}{i}\pnormal^{i}\times(\pbenign-\pnormal)^{n-i}$.
	\smallskip

	Similarly to consistency analysis, we calculated $\ninesofA(CFT)$ and $\ninesofA(\PCBFT)$ for all values of $\ninesnormal$ and $\ninesbenign$ between 1 and 20 in the cases where $t=1$ and $t=2$. Notice that $\pnormal<\pbenign$ is always true, i.e., $\ninesnormal<\ninesbenign$. We observed the following relation for $t=1$:
	
	\smallskip
	
	\[
	\ninesofA(\PCBFT_{t=1})-\ninesofA(CFT_{t=1})=\]
	\[max(2\times\ninesnormal-\ninesbenign,0).
	\]

	When $t=2$, we observed:
	
	\[
	\ninesofA(\PCBFT_{t=2}) = 3\times\ninesnormal-1,
	\]
	
	\[
	\ninesofA(\PCBFT_{t=2})-\ninesofA(CFT_{t=2})=\]
	\[
	\begin{cases}
	3\times\ninesnormal-\ninesbenign,& \ninesbenign<3\times\ninesnormal,\\
	1,& 3\times\ninesnormal\leq\ninesbenign< 4\times\ninesnormal,\\
	0, & \ninesbenign\geq 4\times\ninesnormal.
	\end{cases}
	\]

	\smallskip
	
	\noindent\emph{Example.} When $\pnormal=0.999$ and $\pbenign=0.99999$,
	we have $\ninesofA(\PCBFT_{t=1})-
	\ninesofA(CFT_{t=1})=1$, i.e., $\ninesofA(\PCBFT_{t=1})=5$ and $\ninesofA(CFT_{t=1})=4$. \PCBFT adds 1 nine of availability on top of CFT and achieves 5 nines of availability in total. Besides, \PCBFT adds 2 nines of availability on top of individual machine availability.

	\subsubsection{\PCBFT vs.\ BFT}
	
	From Table~\ref{table:comp}, an asynchronous BFT protocol is available when $n-\lfloor\frac{n-1}{3}\rfloor$ machines are available despite faults of other machines. 
	Thus, $P[\text{BFT is available}]= \sum\limits_{i=n-\lfloor\frac{n-1}{3}\rfloor}^{n}\binom{n}{i}\pnormal^{i}\times(1-\pnormal)^{n-i}$.

	\smallskip
	We calculated $\ninesofA(\PCBFT)$ and $\ninesofA(BFT)$ for all values of $\ninesnormal$ between 1 and 20 in the cases when $t=1$ and $t=2$. In this comparison $\ninesbenign$ does not matter. When $t=1$,
	\[
	\ninesofA(\PCBFT_{t=1})=\ninesofA(BFT_{t=1})=2\times\ninesnormal-1.
	\]
	
	On the other hand, when $t=2$,
	\[
	\ninesofA(\PCBFT_{t=2}) = \ninesofA(BFT_{t=2}) + 1 = 3\times\ninesnormal-1.
	\]
	
	Hence, 
	when $t=1$, \PCBFT has the same number of nines of availability as BFT. When $t=2$, \PCBFT adds 1 nine of availability to BFT.
	
	\section{Related work and concluding remarks}
	\label{sec:conclusion}

	In this paper, we introduced XFT, a novel fault-tolerance model that allows the design of  efficient protocols that tolerate non-crash faults. We demonstrated XFT through \PCBFT, a novel state-machine replication protocol that features many more nines of reliability than the best crash-fault-tolerant (CFT) protocols with roughly the same communication complexity, performance and resource cost. Namely, \PCBFT uses $2t+1$ replicas and provides all the reliability guarantees of CFT, but is also capable of tolerating non-crash faults, as long as a majority of \PCBFT replicas are correct and can communicate synchronously among each other. 
	
	As XFT is entirely realized in software, it is fundamentally different from an established approach that relies on trusted hardware for reducing the resource cost of BFT to $2t+1$ replicas only~\cite{Correia:2004:THL,Levin:2009:TST,Kapitza:2012:CRB:2168836.2168866,VeroneseCBLV13}.
	
	\PCBFT is also different from PASC~\cite{ASC}, which makes CFT protocols tolerate a subset of Byzantine faults using ASC-hardening. ASC-hardening modifies an application by keeping two copies of the state at each replica. It then tolerates Byzantine faults under the ``fault diversity'' assumption, i.e., that a fault will not corrupt both copies of the state in the same way. Unlike 
	\PCBFT, PASC does not tolerate Byzantine faults that affect the entire replica (e.g., both state copies). 
	
	In this paper, we did not explore the impact on varying the number of tolerated faults \emph{per fault class}. In short, this approach, known as the \emph{hybrid} fault model and introduced in \cite{ThambiduraiP88} distinguishes the threshold of non-crash faults (say $b$) despite which safety should be ensured, from the threshold $t$ of faults (of any class) despite which the availability should be ensured (where often $b\le t$). The hybrid fault model and its refinements \cite{Clement:2009:UCS:1629575.1629602,vft} appear orthogonal to our XFT approach. 
	
	Specifically, Visigoth Fault Tolerance (VFT) \cite{vft} is a recent refinement of the hybrid fault model. Besides having different thresholds for non-crash and crash faults, VFT also refines the space between network  synchrony and asynchrony by defining the threshold of network faults that a VFT protocol can tolerate. VFT is, however, different from XFT in that it fixes separate fault thresholds for non-crash and network faults. This difference is fundamental rather than notational, as XFT cannot be expressed by choosing specific values of VFT thresholds. For instance, \PCBFT can tolerate, with $2t+1$ replicas, $t$ partitioned replicas, $t$ non-crash faults and $t$ crash faults, albeit not simultaneously. Specifying such requirements in VFT would yield at least $3t+1$ replicas.  In addition, VFT protocols have more complex communication patterns than \PCBFT. That said, many of the VFT concepts remain orthogonal to XFT. It would be interesting to explore interactions between the hybrid fault model (including its refinements such as VFT) and~XFT in the future.

	Going beyond the research directions outlined above, this paper opens also other avenues for future work. 
	For instance, many important distributed computing problems that build on SMR, such as distributed storage and blockchain, deserve a novel look at them through the XFT prism. 
	
\bibliographystyle{abbrv}
\bibliography{thesis,bibliography,transactional_systems,fault_tolerance,paxos,blockchain}
	
\newpage

\begin{appendices}

\section{\PCBFT example execution}

\begin{figure}[ht!]
	\centering
	\subfloat[without FD]{\includegraphics[scale=0.99]{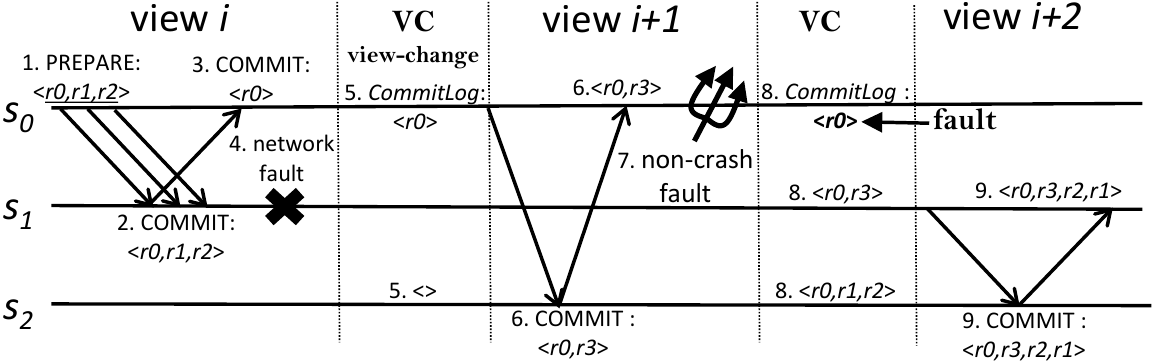}\label{fig:vc_expl}}
	
	\subfloat[with FD]{\includegraphics[scale=0.99]{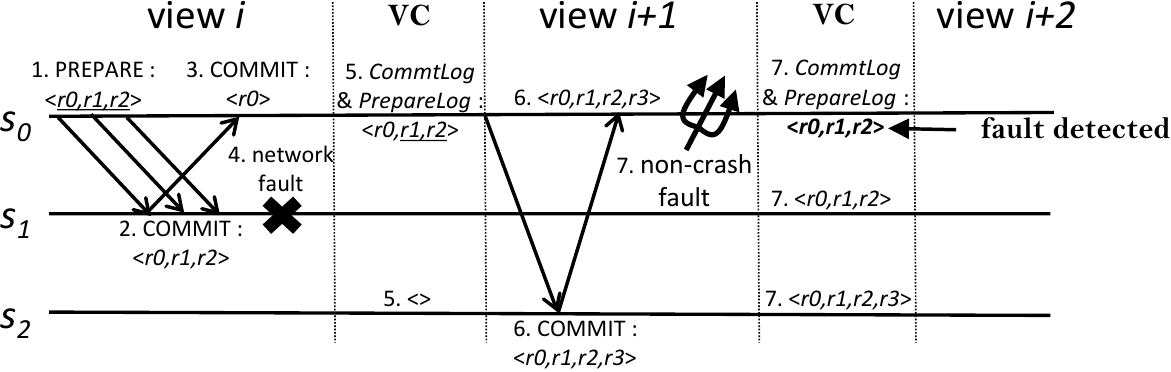}\label{fig:vcfd_expl}}
	\caption{\PCBFT\ example. The view is changed from $i$ to $i+2$, due to the network fault of $s_1$ and the non-crash fault of $s_0$, respectively.}
	\label{fig:vcexample}
\end{figure}

In Fig.~\ref{fig:vcexample} we give an example of \PCBFT execution when $t=1$. The role of each replica in each view is shown in Table~\ref{table:sgcomb}.

In Fig.~\ref{fig:vc_expl}, view change phase proceeds without fault detection. Upon the primary $s_0$ receives requests $r0$, $r1$, and $r2$ from clients, $s_0$ prepares these requests locally and sends \msgtag{commit} messages to the follower $s_1$. Then, $s_1$ commits $r0$, $r1$, and $r2$ locally and sends \msgtag{commit} messages to $s_0$. Because of a network fault, $s_0$ only receives \msgtag{commit} message of $r0$ in a timely manner, thus the view change phase to $i+1$ is activated by $s_0$. During view change to $i+1$, $s_0$ sends the \msgtag{view-change} message with commit log of $r_0$ to all active replicas in view $i+1$ (i.e., $s_0$ and $s_2$). 
In view $i+1$, $r3$ is further committed by $s_0$ and $s_2$. After that, $s_0$ is under non-crash fault and the view is changed to $i+2$. During view change to $i+2$, $s_1$ and $s_2$ provide all their commit logs to new active replicas (i.e., $s_1$ and $s_2$), whereas non-crash faulty replica $s_0$ only reports the commit log of $r0$. Outside anarchy, requests $r0$ and $r3$ are committed in new view $i+2$ by receiving the \msgtag{view-change} message from $s_2$. Request $r_3$ is also committed by receiving the \msgtag{view-change} message from $s_1$. In view $i+2$, $r1$ is finally committed by every active replica. 

In example of Figure~\ref{fig:vcfd_expl}, \PCBFT fault detection is enabled. In view $i$, the execution is the same as in Figure~\ref{fig:vc_expl}. During view change to $i+1$, commit log of $r0$ and prepare logs of $r1$ and $r2$ are sent by $s_0$, which are committed by $s_0$ and $s_2$ in view $i+1$, as well as the new request $r3$. The same as before, $s_0$ is non-crash faulty and the view is changed to $i+2$. During view change to $i+2$, commit logs of $r0$, $r1$, $r2$ and $r3$ are sent by $s_2$. At the same time, because of missing prepare log of $r3$, the fault of $s_0$ is detected with the help of the \msgtag{view-change} message from $s_2$.

\label{apd:exp}

\clearpage

\section{\PCBFT pseudocode}
\label{apd:pseudocode}
\newcommand{\hmt}[1]{/* #1 */}

In this appendix we give the pseudocode of \PCBFT.	For simplicity reason, we assume that signature/MAC attached to each message always correctly verifies. Figure~\ref{fig:message} gives the definition of message fields and local variables for all components of \PCBFT. Readers can refer to Section~\ref{sec:nutshell} for protocol description. 
	
This appendix is organized incrementally as follows. Section~\ref{sec:pcc} gives the pseudocode of \PCBFT common case. Section~\ref{sec:vc_detail} gives the pseudocode of the view change mechanism. Section~\ref{sec:liveness} describes and gives the pseudocode of clients' request retransmission mechanism that deals with faulty primary. Finally, Section~\ref{sec:vcfd_pcode} depicts the modification to the view change protocol to enable Fault Detection and gives the pseudocode.

\newcommand{\cbox}[1] {\vspace*{0.2cm} \noindent
\hspace*{0cm}\fbox{
\begin{minipage}{0.95\textwidth}
{\sf #1}
\end{minipage}
}\\}

\begin{figure}[!ht]
	\begin{center}
		\cbox{
			\small{
				Common case :\\
				$c,op,ts_c$ - id of the client, operation, client timestamp\\
				$req_c$ - ongoing request at client $c$\\
				$n$ - total number of replicas\\
				$\Pi$ - set of $n$ replicas\\
				$i$ - current view number\\
				$s_j$ - replica id\\
				$\vs_{i}$ - set of $t+1$ replicas in synchronous group in view $i$\\
				$ps_i$ - the primary in view $i$ ($ps_i\in\vs_i$)\\
				$fs_i$ - the follower in view $i$ for $t=1$ ($fs_i\in\vs_i$)\\
				$fs^k_i$ - the followers in view $i$ for $t\geq 2$ ($fs^k_i\in\vs_i$)\\
				$req$ - client request\\
				$rep$ - reply of client request\\
				$sn_{s_j}$ - sequence number prepared at replica $s_j$\\
				$ex_{s_j}$ - sequence number executed at replica $s_j$\\
				$D(m)$ - digest of a message $m$\\
				$PrepareLog_{s_j}$ - array of prepared proof at replica $s_j$\\
				$CommitLog_{s_j}$ - array of commit proofs at replica $s_j$\\
				\newline
				View change :\\
				$SusSet_{s_j}$ - set of \msgtag{suspect} messages cached for view-change at replica $s_j$\\
				$timer^{net}_i$ - network establishment timer for view $i$\\
				$\Delta$ - maximum message delay between two correct replicas, beyond which a network fault is declared\\
				$timer^{vc}_i$ - view-change timer in view change to $i$\\
				$VCSet_{s_j}^{i}$ - set of \msgtag{view-change} messages collected in view change to $i$ at replica $s_j$\\
				$CommitLog_{s_j}^{i}$ - array of most recent commit proofs selected from $VCSet_{s_j}^{i}$ at replica $s_j$\\
				$End(log)$ - end index of array $log$\\
				\newline
				Fault detection :\\
				$FinalProof_{s_j}$ - array of $t+1$ \msgtag{vc-confirm} messages which prove that $\forall s_k\in\vs_{i}$ collected the same $VCSet_{s_k}^{i}$\\
				$pre_{s_j}$ - the view number in which $PrepareLog_{s_j}$ is generated\\ 
				$FinalSet_{s_j}^{i}$ - set of $t+1$ \msgtag{vc-final} messages collected in view change to $i$ at replica $s_j$\\
				$PrepareLog_{s_j}^{i}$ - array of most recent prepare proof selected from $VCSet_{s_j}^{i}$ at replica $s_j$
			}}
		\caption{\PCBFT\ common case: Message fields and local variables.} \label{fig:message}
		\end{center}
	\end{figure}	
	
	\clearpage
	
	\subsection{Common case}
	\label{sec:pcc}
	
	In common case, we assume that all replicas are in the same view. Algorithm~\ref{alg:f1} and Algorithm~\ref{alg:f2} describe the common case protocol when $t=1$ and $t\geq 2$, respectively. Figure~\ref{fig:Atlas} gives the message pattern.

	\def\NoNumber#1{{\def\alglinenumber##1{}\State #1}\addtocounter{ALG@line}{-1}}
	
	\begin{algorithm}
		\caption{Common case when $t=1$.}\label{alg:f1}
		\begin{minipage}{0.95\textwidth}
		\small
		\begin{algorithmic}[1]
			\Statex Initialization:
			\Statex client : $ts_c\leftarrow 0$; $req_c\leftarrow nil$
			\Statex replica : $sn_{s_j}\leftarrow 0; ex_{s_j}\leftarrow 0; PrepareLog_{s_j}=[]; CommitLog_{s_j}=[]$
			\newline
			\State \textbf{upon} invocation of $propose(op)$ at client $c$ \textbf{do}
			\State \indent \textbf{inc}($ts_c$)
			\State \indent send $req_c\leftarrow\angular{\interface{\ndR}, op,ts_c,c}_{\sigma_c}$ to the primary $ps_i\in\vs_i$
			\State \indent start $timer_{c}$
			\newline
			\State \textbf{upon} reception of $req=\angular{\interface{\ndR}, op,ts,c}_{\sigma_c}$ from client $c$ at $ps_i$ \textbf{do} \cmt{primary} 
			\State \indent \textbf{inc}($sn_{ps_i}$)
			\State \indent $m_{ps_i}\leftarrow\angular{\MSGCOMMIT, D(req), sn_{ps_i}, i}_{\sigma_{ps_i}}$
			\State \indent $PrepareLog_{ps_i}[sn_{ps_i}]\leftarrow\angular{req,m_{ps_i}}$ 
			\State \indent send $\angular{\req,m_{ps_i}}$ to the follower $fs_i$
			\newline
			\State \textbf{upon} reception of $\angular{\req,m_{ps_i}=\angular{\MSGCOMMIT, d_{req}, sn, i}_{\sigma_{ps_i}}}$ from the primary $ps_i$ at $fs_i$ \textbf{do} \cmt{follower}
			\State \indent \textbf{if} $sn=sn_{fs_i}+1$ \textbf{and} $D(req)=d_{req}$ \textbf{then}
			\State \indent\indent \textbf{inc}($sn_{fs_i}$)
			\State \indent\indent $rep\leftarrow$ execute $req$
			\State \indent\indent \textbf{inc}($ex_{fs_i}$)
			\State \indent\indent $m_{fs_i}\leftarrow\angular{\MSGCOMMIT, D(req), sn, i, req.ts_c, D(rep)}_{\sigma_{fs_i}}$
			\State \indent\indent $CommitLog_{fs_i}[sn]\leftarrow\angular{req,m_{ps_i},m_{fs_i}}$ 
			\State \indent\indent send $m_{fs_i}$ to the primary $ps_i$
			\newline
			\State \textbf{upon} reception of $m_{fs_i}=\angular{\MSGCOMMIT, d_{req}, sn, i, ts, d_{rep}}_{\sigma_{fs_i}}$ from the follower $fs_i$ at $ps_i$ \textbf{do}
			\State \indent \textbf{if} $D(PrepareLog_{ps_i}[sn].req)=d_{req}$ \textbf{then}
			\State \indent\indent $CommitLog_{ps_i}[sn]\leftarrow\angular{req,m_{ps_i},m_{fs_i}}$
			\newline
			\State \textbf{upon} $CommitLog_{ps_i}[ex_{ps_i}+1]\neq nil$ at $ps_i$ \textbf{do}
			\State \indent \textbf{inc}($ex_{ps_i}$)
			\State \indent $rep\leftarrow$ execute $CommitLog_{ps_i}[ex_{ps_i}].req$
			\State \indent \textbf{if} $D(rep)=CommitLog_{ps_i}[ex_{ps_i}].m_{fs_i}.d_{rep}$ \textbf{then}
			\State \indent\indent send $\angular{\angular{\MSGREPLY, sn, i, ts, rep}_{\mu_{ps_i,c}}, m_{fs_i}}$ to $CommitLog_{ps_i}[ex_{ps_i}].req.c$
			\newline
			\State \textbf{upon} reception of $\angular{r_{ps_i},m_{fs_i}}$ from the primary $ps_i$ at client $c$, where
			\indent \NoNumber{$r_{ps_i}=\angular{\MSGREPLY, sn, i, ts, rep}_{\mu_{ps_i,c}}$}
			\indent \NoNumber{$m_{fs_i}=\angular{\MSGCOMMIT, d'_{req}, sn', i', ts', d_{rep}}_{\sigma_{fs_i}}$} \textbf{do}
			\State \indent \textbf{if} $sn=sn'$ \textbf{and} $i=i'$ \textbf{and} $ts=ts'=req.ts_c$ \textbf{and} $D(rep)=d_{rep}$ \textbf{then}
			\State \indent\indent deliver $rep$
			\State \indent\indent stop $timer_c$
		\end{algorithmic}
		\end{minipage}
	\end{algorithm}
	
	\clearpage
	
	\begin{algorithm}
		\caption{Common case when $t>1$.}\label{alg:f2}
		\small
		\begin{minipage}{0.95\textwidth}
		\begin{algorithmic}[1]
			\Statex Initialization:
			\Statex client : $ts_c\leftarrow 0$; $req_c\leftarrow nil$
			\Statex replica : $sn_{s_j}\leftarrow 0; ex_{s_j}\leftarrow 0; PrepareLog_{s_j}=[]; CommitLog_{s_j}=[]$
			\newline
			\State \textbf{upon} invocation of $propose(op)$ at client $c$ \textbf{do}
			\State \indent \textbf{inc}($ts_c$)
			\State \indent send $req_c\leftarrow\angular{\interface{\ndR}, op,ts_c,c}_{\sigma_c}$ to the primary $ps_i\in\vs_i$
			\State \indent start $timer_{c}$
			\newline
			\State \textbf{upon} reception of $req=\angular{\interface{\ndR},op,ts,c}_{\sigma_c}$ from client $c$ at $ps_i$ \textbf{do} \cmt{primary}
			\State \indent \textbf{inc}$(sn_{ps_i})$
			\State \indent $m_{ps_i}\leftarrow\angular{\MSGPREPARE,D(req),sn_{ps_i},i}_{\sigma_{ps_i}}$
			\State \indent $PrepareLog_{ps_i}[sn]\leftarrow\angular{\req,m_{ps_i}}$
			\State \indent send $\angular{\req,m_{ps_i}}$ to $fs^k_i\in\sg_i$
			\newline
			\State \textbf{upon} reception of $\angular{\req,m_{ps_i}=\angular{\MSGPREPARE, d_{req}, sn, i}_{\sigma_{ps_i}}}$ from the primary $ps_i$ at $fs^k_i$ \textbf{do} \cmt{follower}
			\State \indent \textbf{if} $sn=sn_{fs^k_i}+1$ \textbf{and} $D(\req)=d_{req}$ \textbf{then}
			\State \indent\indent \textbf{inc}$(sn_{fs^k_i})$
			\State \indent\indent	$PrepareLog_{fs^k_i}[sn]\leftarrow\angular{req,m_{ps_i}}$
			\State \indent\indent $m_{fs^k_i}\leftarrow\angular{\MSGCOMMIT, D(req), sn, i, fs^k_i}_{\sigma_{fs^k_i}}$
			\State \indent\indent send $m_{fs^k_i}$ to $\forall s_k\in\vs_i$
			\newline
			\State \textbf{upon} reception of $m_{fs^k_i}=\angular{\MSGCOMMIT, d_{req}, sn, i, fs^k_i}_{\sigma_{fs^k_i}}$ from every follower $fs^k_i\in\vs_i$ at $s_j\in\vs_i$ \textbf{do}
			\State \indent $CommitLog_{s_j}[sn]\leftarrow\angular{req,m_{ps_i},m_{fs^1_i}...m_{fs^f_i}}$
			\newline
			\State \textbf{upon} $CommitLog_{s_j}[ex_{s_j}+1]\neq nil$ at $s_j$ \textbf{do}
			\State \indent \textbf{inc}$(ex_{s_j})$
			\State \indent $rep\leftarrow$ execute $CommitLog_{s_j}[ex_{s_j}].req$
			\State \indent send $\angular{\MSGREPLY, sn, i, req.ts_c, rep}_{\mu_{s_j,c}}$ to client $c$, where $c=CommitLog_{s_j}[ex_{s_j}].req.c$)
			\newline
			\State \textbf{upon} reception of $t+1$ $\MSGREPLY$ messages $\angular{\MSGREPLY, sn, i, ts, rep}_{\mu_{s_j,c}}$ at client $c$ \textbf{do}
			\State \indent \textbf{if} $t+1$ $\MSGREPLY$ messages are with the same $sn$, $i$, $ts$ and $rep$ \textbf{and} $ts=req.ts_c$ \textbf{then}
			\State \indent\indent deliver $rep$
			\State \indent\indent stop $timer_c$
		\end{algorithmic}
	\end{minipage}
	\end{algorithm}

	\clearpage
	
	\subsection{View-change}
	\label{sec:vc_detail}

	\newcommand{\stablecheckpoint}{\ensuremath{stb\_ck}}
	\newcommand{\commitset}{\ensuremath{cmt\_set}}
	\newcommand{\favaritelist}{\ensuremath{favorite\_list}}
	\newcommand{\candidatenumber}{\ensuremath{candidate\_number}}
	
	\newcommand{\leaderchangeset}{\ensuremath{lc\_set}}
	\newcommand{\pf}{\ensuremath{proof}}

	The message pattern of view-change w/o fault detection is given in Figure~\ref{fig:VRC}. Algorithm~\ref{alg:vc} shows the corresponding pseudocode. The description of view change can be found in Section~\ref{sec:VC}.

	\begin{algorithm}
		\caption{View change at replica $s_j$.}\label{alg:vc}
		\small
		\begin{minipage}{0.95\textwidth}
		\begin{algorithmic}[1]
			\Statex Initialization:
			\Statex $SusSet_{s_j}\leftarrow \emptyset; VCSet_{s_j}^i\leftarrow \emptyset; CommitLog_{s_j}^i\leftarrow []$
			\newline
			\State \textbf{upon} suspicion of view $i$ \textbf{and} $s_j\in\vs_i$ \textbf{do} 
			\State \indent send $\angular{\msgtag{suspect}, i, s_j}_{\sigma_{s_j}}$ to $\forall s_k\in \Pi$
			\newline
			\State \textbf{upon} reception of $m=\angular{\msgtag{suspect}, i', s_k}_{\sigma_{s_k}}$ \textbf{and} $s_k\in\vs_{i'}$ \textbf{do}
			\State \indent $SusSet_{s_j}\leftarrow SusSet_{s_j}\cup \{m\}$
			\State \indent forward $m$ to $\forall s_k\in \Pi$
			\newline
			\State \textbf{upon} $\exists \angular{\msgtag{suspect}, i, s_k}_{\sigma_{s_k}}\in SusSet_{s_j}$ \textbf{do} \cmt{enter each view in order}
			\State \indent \textbf{inc}($i$) (i.e., ignore any message in preceding view)
			\State \indent send
			\begingroup
			$\angular{\msgtag{view-change},i,s_j,CommitLog_{s_j}}_{\sigma_{s_j}}$
			\endgroup
			to $\forall s_k\in \vs_{i}$
			\State \indent \textbf{if} $s_j\in\vs_{i}$ \textbf{then}
			\State \indent\indent start $timer^{net}_{i}\leftarrow 2\Delta$
			\newline
			\State \textbf{upon} reception of $m=\angular{\msgtag{view-change},i,s_k,CommitLog}_{\sigma_{s_k}}$ from replica $s_k$ \textbf{do}
			\State \indent $VCSet_{s_j}^{i}\leftarrow VCSet_{s_j}^{i}\cup\{m\}$
			\newline
			\State \textbf{upon} $|VCSet_{s_j}^{i}|=n$ \textbf{or} (expiration of $timer^{net}_{i}$ \textbf{and} $|VCSet_{s_j}^{i}|\geq n-t$) \textbf{do} 
			\State \indent send $\angular{\msgtag{vc-final},i,s_j,VCSet_{s_j}^{i}}_{\sigma_{s_j}}$ to $\forall s_{k}\in\vs_{i}$
			\State \indent start $timer^{vc}_{i}$
			\newline
			\State \textbf{upon} reception of $m_*=\angular{\msgtag{vc-final},i,s_k,VCSet}_{\sigma_{s_k}}$ from every $s_k\in\vs_{i}$ \textbf{do}
			\State \indent $VCSet_{s_j}^{i}\leftarrow VCSet_{s_j}^{i}\cup \{\forall m:m\in VCSet$ in any $m_k$\} 
			\State \indent \textbf{for} $sn : 1..End(\forall CommitLog | \exists m\in VCSet_{s_j}^{i}$ : $CommitLog$ is in $m$) 
			\textbf{do} 
			\State \indent\indent $CommitLog_{s_j}^{i}[sn]\leftarrow CommitLog[sn]$ with the \emph{highest} view number
			\State \indent \textbf{if}  $s_j=ps_{i}$ \textbf{then} \cmt{primary}
			\State \indent\indent \textbf{for} $sn:1..End(CommitLog_{s_j}^{i})$ \textbf{do}
			\State \indent\indent\indent $req\leftarrow CommitLog_{s_j}^{i}[sn].req$ 
			\State \indent\indent\indent $PrepareLog[sn]\leftarrow$ $\angular{req,\angular{\MSGPREPARE,D(req),sn,i}_{\sigma_{ps_{i}}}}$
			\State \indent\indent send $\angular{\msgtag{new-view},i,PrepareLog}_{\sigma_{ps_{i}}}$ to $\forall s_k\in\vs_{i}$
			\newline
			\State \textbf{upon} reception of $\angular{\msgtag{new-view},i,PrepareLog}_{\sigma_{ps_{i}}}$ from the primary $ps_{i}$ \textbf{do}
			\State \indent \textbf{if} $PrepareLog$ is matching with $CommitLog_{s_j}^{i}$ \textbf{then}
			\State \indent\indent $PrepareLog_{s_j}\leftarrow PrepareLog$
			\State \indent\indent reply and process $\forall m\in PrepareLog$ as in common case
			\State \indent\indent $sn_{s_j}\leftarrow End(PrepareLog)$
			\State \indent\indent $ex_{s_j}\leftarrow End(PrepareLog)$
			\State \indent\indent stop $timer^{vc}_{i}$
			\State \indent \textbf{else}
			\State \indent\indent suspect view $i$
			\newline
			\State \textbf{upon} expiration of $timer^{vc}_{i}$ \textbf{do}
			\State \indent suspect view $i$
		\end{algorithmic}
		\end{minipage}
	\end{algorithm}

\clearpage

\subsection{Request retransmission}
\label{sec:liveness}

In order to provide availability with respect to faulty primary or followers, as well as long-lived network faults within the synchronous group, we propose a request retransmission mechanism which broadcasts the request to all active replicas upon retransmission timer expires at client side. Retransmission mechanism requires every active replica to monitor the progress. In case a request is not executed and replied in a timely manner, the correct active replica in the synchronous group will eventually suspect the view.
	
More specifically (the pseudocode is given in Algorithm~\ref{alg:liveness}), if a client $c$ does not receive the matching replies of request $req_c$ in a timely manner, $c$ re-sends $req_c$ to all active replicas in current view $i$ by \angular{\msgtag{re-send},$req_c$}. Any active replica $s_j\in\vs_i$, upon receiving \angular{\msgtag{re-send},$req_c$} from $c$, (1) forwards $req_c$ to the primary $ps_i\in\vs_i$ if $s_j\neq ps_i$, (2) starts a timer $timer_{req_c}$ locally, and (3) asks each active replica to sign the reply. Upon $timer_{req_c}$ expires and the active replica $s_j\in\vs_i$ has not received $t+1$ signed replies, $s_j$ suspects view $i$ and sends the \msgtag{suspect} message to the client $c$; otherwise, $s_j$ forwards $t+1$ signed replies to client $c$. 
	
Upon receiving \msgtag{suspect} message $m$ for view $i$, client $c$ forwards $m$ to every active replica in view $i+1$. This step serves to guarantee that the view-change can actually happen at all correct replicas. Then client $c$ forwards $req_c$ to the primary of view $i+1$.

	\begin{algorithm}
		\caption{Client request retransmission.}\label{alg:liveness}
		\small
		\begin{minipage}{0.95\textwidth}
		\begin{algorithmic}[1]
			\State \textbf{upon} expiration of $timer_c$ at client $c$ \textbf{do}
			\State \indent send \angular{\msgtag{re-send},$req_c$} to $\forall s_j\in\vs_i$
			\newline
			\State \textbf{upon} reception of \angular{\msgtag{re-send},$req_c$} at $s_j\in\vs_i$ \textbf{do}
			\State \indent \textbf{if} $s_j\neq ps_i$ \textbf{then}
			\State \indent\indent send $req_c$ to $ps_i\in\vs_i$ 
			\State \indent start $timer_{req_c}$
			\State \indent ask $\forall s_j\in\vs_i$ to sign the reply of $req_c$
			\newline
			\State \textbf{upon} expiration of $timer_{req_c}$ at replica $s_j\in\vs_i$ \textbf{do}
			\State \indent suspect view $i$
			\State \indent send $\angular{\msgtag{suspect}, i, s_j}_{\sigma_{s_j}}$ to client $c$
			\newline
			\State \textbf{upon} reception of $m=\angular{\msgtag{suspect}, i, s_k}_{\sigma_{s_k}}$ at client $c$ \textbf{and} $s_k\in\vs_i$ \textbf{and} $c$ is in view $i$ \textbf{do}
			\State \indent enter view $i+1$
			\State \indent send $m$ to $\forall s_j\in\vs_{i+1}$
			\State \indent send $req_c$ to $ps_{i+1}$
			\State \indent start $timer_c$
			\newline
			\State \textbf{upon} execution of $req_c$ at $s_j$\footnote{by line 13 or 23 in Algorithm~\ref{alg:f1} or line 20 in Algorithm~\ref{alg:f2}} \textbf{do}
			\Statex \indent\indent \hmt{sign the reply by each active replica}
			\State \indent send $\angular{\MSGREPLY,sn,i,req.ts_c,rep}_{\sigma_{s_j}}$ to $\forall s_j\in\vs_i$ 
			\newline
			\State \textbf{upon} reception of $m_k=\angular{\MSGREPLY,sn,i,ts,rep}_{\sigma_{s_k}}$ from every $s_k\in\vs_i$ at replica $s_j$ \textbf{do} \Statex \cmt{collect $t+1$ signed replies}
			\State \indent \textbf{if} $m_1,m_2,...,m_{t+1}$ are with the same $sn$, $i$, $ts$ and $rep$ \textbf{then}
			\State \indent\indent $replies\leftarrow$ \{$m_1,m_2,...,m_{t+1}$\}
			\State \indent\indent send $\angular{\msgtag{signed-reply},replies}$ to client $c$
			\State \indent\indent stop $timer_{req_c}$
		\end{algorithmic}
		\end{minipage}
	\end{algorithm}

\clearpage

\subsection{Fault detection}
\label{sec:vcfd_pcode}
	
	In this section we describe \PCBFT with Fault Detection (FD). Specifically, in order to detect all the fatal faults that can possibly violate consistency in anarchy, view change to $i+1$ with FD includes the following modifications.

\begin{figure}
\centering
\includegraphics[scale=0.90]{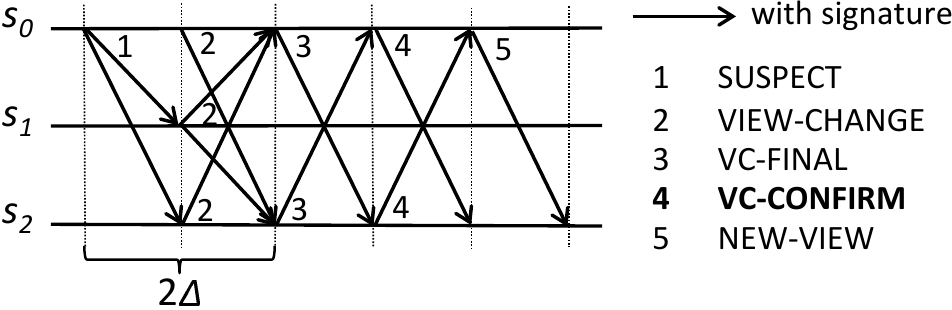}
\caption{Message pattern of \PCBFT\ view-change with fault detection: \msgtag{vc-confirm} phase is added; synchronous group is changed from ($s_0$,$s_1$) to ($s_0$,$s_2$).}
\label{fig:vcfd}
\end{figure}

\begin{itemize}
		\item Every replica $s_j$ appends its prepare logs $PrepareLog_{s_j}$ into the \msgtag{view-change} message when replying to active replicas in view $i+1$. Besides, synchronous group $\vs_{i+1}$ prepares and commits requests piggybacked in commit \emph{or} prepare logs. The selection rule is almost the same as in view change without FD: for each sequence number $sn$, the request with the highest view number $i'\leq i$ is selected, either in a commit log or in a prepare log.
		
		\item \PCBFT FD additionally inserts a \msgtag{vc-confirm} phase after exchanging \msgtag{view-change} messages among active replicas in view $i+1$, i.e., after receiving $t+1$ \msgtag{vc-final} messages (see Figure~\ref{fig:VRC} and Figure~\ref{fig:vcfd} for the comparison). In \msgtag{vc-confirm} phase, every active replica $s_j\in\vs_{i+1}$ (1) detects potential faults in the \msgtag{view-change} messages in $VCSet^{i+1}_{s_j}$ and adds the faulty replica to set $FSet$; (2) removes faulty messages from $VCSet^{i+1}_{s_j}$; and, (3) signs and sends $\msg{\msgtag{vc-confirm},i+1,D(VCSet^{i+1}_{s_j})}_{\sigma_{s_j}}$  to every active replica in $\vs_{i+1}$. Upon $s_j\in\vs_{i+1}$ receives $t+1$ \msgtag{vc-confirm} messages with matching $D(VCSet^{i+1}_*)$, $s_j$ (1) inserts the \msgtag{vc-confirm} messages into set $FinalProof_{s_j}[i+1]$; and (2) prepares and commits the requests selected based on $VCSet^{i+1}_{s_j}$. $FinalProof_{s_j}[i+1]$ serves to prove that $t+1$ active replicas in $\vs_{i}$ have agreed on the set of \emph{filtered} \msgtag{view-change} messages. 
	
		\item Every replica $s_j$ appends $FinalProof_{s_j}[i']$ into the \msgtag{view-change} message when replying to active replicas in new view, where $i'$ is the view in which $PrepareLog_{s_j}$ is generated. In case a prepare log in $PrepareLog_{s_j}$ is not consistent with some commit log, $FinalProof_{s_j}[i]$ can prove that there exists correct replica $s_j\in\vs_{i'}$ which can prove the fault of the prepare log.
	\end{itemize}

		Algorithm~\ref{alg:vcfd} gives the modifications based on Algorithm~\ref{alg:vc} for \PCBFT with fault detection mechanism. Algorithm~\ref{alg:fd} enumerates all types of faults that can and must be detected by correct active replicas. Figure~\ref{fig:vcfd} gives the new message pattern.

	\begin{algorithm}
		\caption{Modifications for fault detection at replica $s_j$.}\label{alg:vcfd}
		\small
		\begin{minipage}{0.95\textwidth}
		\begin{algorithmic}[1]
			\Statex Initialization:
			\Statex $FinalProof_{s_j}\leftarrow []; pre_{s_j}\leftarrow 0; FinalSet_{s_j}^{i}\leftarrow \emptyset; PrepareLog_{s_j}^{i}\leftarrow []; FSet\leftarrow []$
			\newline
			\Statex /* replace line 8 in Algorithm~\ref{alg:vc} by : */
			\State \indent send $m=\angular{\msgtag{view-change},i,s_j,CommitLog_{s_j},$ $PrepareLog_{s_j},FinalProof_{s_j}[pre_{s_j}]}_{\sigma_{s_j}}$ to $\forall s_k\in\vs_{i}$
			\newline
			\Statex /* replace line 11 in Algorithm~\ref{alg:vc} by : */
			\State \textbf{upon} reception of $m=\angular{\msgtag{view-change},i,s_k,CommitLog,$ $PrepareLog,FinalProof}_{\sigma_{s_k}}$ from replica $s_k$ \textbf{do}
			\newline
			\Statex /* replace lines $18\scriptsize{\sim} 24$ in Algorithm~\ref{alg:vc} by : */
			\State \indent \textsc{faultDetection($vcSet_{s_j}^{i}$)} \cmt{refer to Algorithm~\ref{alg:fd}}
			\State \indent \textbf{for} $\forall m: m\in vcSet_{s_j}^{i}$ \textbf{and} $m$ from replica $s\in FSet$ \textbf{do}
			\State \indent\indent remove $m$ from $vcSet_{s_j}^{i}$  
			\State \indent send $\angular{\msgtag{vc-confirm},i,D(vcSet_{s_j}^{i})}_{\sigma_{s_j}}$ to $\forall s_k\in\vs_{i}$
			\newline
			\Statex /* new event handler */
			\State \textbf{upon} reception of $m_*=\angular{\msgtag{vc-confirm},i,d_{vcSet}}_{\sigma_{s_k}}$ from every $s_k\in\vs_{i}$ \textbf{do}
			\State \indent \textbf{if} $m_1,m_2,...m_{f+1}$ are \emph{not} with the same $d_{vcSet}$ \textbf{then}
			\State \indent\indent suspect view $i$
			\State \indent\indent \textbf{return}
			\State \indent $FinalProof_{s_j}[i]\leftarrow$ \{$m_1,m_2,...m_{f+1}$\}
			\State \indent \textbf{for} $sn : 1..End(\forall CommitLog | \exists m\in VCSet_{s_j}^{i}$ : $CommitLog$ is in $m$) \textbf{do}
			\State \indent\indent $CommitLog_{s_j}^{i}[sn]\leftarrow CommitLog[sn]$ with the \emph{highest} view number
			\State \indent \textbf{for} $sn : 1..End(\forall PrepareLog | \exists m\in VCSet_{s_j}^{i}$ : $PrepareLog$ is in $m$) \textbf{do}
			\State \indent\indent $PrepareLog_{s_j}^{i}[sn]\leftarrow PrepareLog[sn]$ with the \emph{highest} view number
			\State \indent \textbf{if}  $s_j=ps_{i}$ \textbf{then} \cmt{primary} 
			\State \indent\indent \textbf{for} $sn:1..End(PrepareLog_{s_j}^{i}|CommitLog_{s_j}^{i})$ \textbf{do}
			\State \indent\indent\indent $req\leftarrow CommitLog_{s_j}^{i}[sn].\req$
			\State \indent\indent\indent \textbf{if} $req=null$ \textbf{or} $PrepareLog_{s_j}^{i}[sn]$ is generated in a higher view than $CommitLog_{s_j}^{i}[sn]$ \textbf{then}
			\State \indent\indent\indent\indent $req\leftarrow PrepareLog_{s_j}^{i}[sn].\req$
			\State \indent\indent\indent $PrepareLog[sn]\leftarrow$ $\angular{req,\angular{\MSGPREPARE,D(req),sn,i}_{\sigma_{s_j}}}$
			\State \indent\indent send $\angular{\msgtag{new-view},i,PrepareLog}_{\sigma_{s_j}}$ to $\forall s_k\in\vs_{i}$
			\newline
			\Statex /* replace line 26 in Algorithm~\ref{alg:vc} by : */
			\State \indent \textbf{if} $PrepareLog$ is matching with $CommitLog_{s_j}^{i}$ and $PrepareLog_{s_j}^{i}$ \textbf{then}
			\newline
			\Statex /* add this command after line 27 in Algorithm~\ref{alg:vc} : */
			\State \indent\indent $pre_{s_j}\leftarrow i$ \cmt{update the view in which $PrepareLog_{s_j}$ is generated}
		\end{algorithmic}
		\end{minipage}
	\end{algorithm}

	\begin{algorithm}
		\caption{Fault detection function at replica $s_j$.}\label{alg:fd}
		\small
		\begin{minipage}{0.95\textwidth}
		\begin{algorithmic}[1]
			\State \textbf{function} \textsc{faultDetection($VCSet$)}
			
			\State \indent $\forall sn$ and $m,m'\in VCSet$ from replicas $s_k$ and $s_{k'}$, respectively,
			\newline
			\State \indent\indent (state loss) \textbf{if} $s_k,s_{k'}\in\vs_{i'}$ ($i'<i$) \textbf{and} $CommitLog'[sn]$ in $m'$ is generated in view $i'$ \textbf{and} $PrepareLog$ is in $m$ \textbf{and} $PrepareLog[sn]=nil$ \textbf{then} ($s_k$ is faulty) 
			\State \indent\indent\indent send $\angular{\msgtag{state-loss},i,s_k,sn,m,m'}$ to $\forall s_{k''}\in \Pi$
			\State \indent\indent\indent add $s_k$ to $FSet$			
			\newline
			\State \indent\indent (fork-I) \textbf{if} $s_k,s_{k'}\in\vs_{i'}$ ($i'< i$) \textbf{and} $PrepareLog[sn]$ in $m$ is generated in view $i''$ \textbf{and} $CommitLog'[sn]$ in $m'$ is generated in view $i'$ \textbf{and} (($i''=i'$ \textbf{and} $PrepareLog[sn].req\neq CommitLog'[sn].req$) \textbf{or} $i''<i$) \textbf{then} ($s_k$ is faulty) 
			\State \indent\indent\indent send $\angular{\msgtag{fork-i},i,s_k,sn,m,m'}$ to $\forall s_{k''}\in \Pi$
			\State \indent\indent\indent add $s_k$ to $FSet$
			\newline
			\State \indent\indent (fork-II-query) \textbf{if} $PrepareLog[sn]$ in $m$ is generated in view $i''$ ($i''< i$) \textbf{and} $CommitLog'[sn]$ in $m'$ is generated in view $i'$ ($i'< i''< i$) \textbf{and} ($PrepareLog[sn]=null$ \textbf{or} $PrepareLog[sn].req\neq CommitLog'[sn].req$) \textbf{then} ($s_k$ \emph{might be} faulty) 
			\State \indent\indent\indent send $\angular{\msgtag{fork-ii-query},i,s_k,sn,m}$ to $\forall s_{k''}\in\vs_{i''}$
			\State \indent\indent\indent wait for $2\Delta$ time
	\newline
	\State \textbf{upon} reception of $\angular{\msgtag{fork-ii-query},i,s_k,sn,m}$ at $s_j$, where $finalProof$ in $m$ is generated in view $i''$ and $s_j\in\vs_{i''}$ \textbf{do}
			\State \indent \textbf{if} $PrepareLog[sn]$ in $m$ is not consistent with $VCSet_{s_j}^{i''}$ \textbf{then} 
			\State \indent\indent send $\angular{\msgtag{fork-ii},i,s_k,sn,m,finalProof_{s_j}[i''], finalSet^{i''}_{s_j}}$ to $\forall s_k\in \Pi$ 
			\newline
	\State \textbf{upon} reception of  $\angular{\msgtag{fork-ii},i,s_k,sn,m,finalProof, finalSet}$ \textbf{do}
	\State \indent add $s_k$ to $FSet$
	\newline
			\State \textbf{upon} reception of \msgtag{state-loss}, \msgtag{fork-i} or \msgtag{fork-ii} message $m$ \textbf{do}
			\State \indent forward $m$ to $\forall s_k\in \Pi$
	\end{algorithmic}
	\end{minipage}
	\end{algorithm}

\clearpage

\section{\PCBFT correctness proof}
\label{apd:proof}
\newcommand{\st}{\ensuremath{st}\xspace}
\newcommand{\timer}{\ensuremath{timer}\xspace}

\newcommand{\artproof}[1]{\textbf{\textit{Proof}}: #1}
\newcommand{\tlproof}[2]{\textbf{\textsc{Proof}} : #1\begin{enumerate}#2\qed\end{enumerate}}

\newcommand{\tlproofs}[1]{\textbf{\textsc{Proof}}: #1\qed\\}

In this appendix, we first prove safety (consistency) and liveness (availability) properties of \PCBFT. To prove safety (Section~\ref{sec:safety}), we show that when \PCBFT\ is outside anarchy, consistency is guaranteed. In liveness section (Section~\ref{sec:prliveness}), we show that \PCBFT\ can make progress with at most $t$ faulty replicas and any number of faulty clients, if eventually the system is synchronous (i.e., eventual synchrony). 

Then, in Section~\ref{sec:prfaultdetection}, we prove that the fault detection mechanism is strong completeness and strong accuracy outside anarchy, with respect to non-crash faults which can violate consistency in anarchy.

We use the notation in Figure~\ref{fig:proof} to facilitate our proof of \PCBFT. All predicates in Figure~\ref{fig:proof} are defined with respect to \emph{benign} clients and replicas.

\newcommand{\clientDelivered}[1]{delivered(\ensuremath{#1})}
\newcommand{\before}[1]{before(\ensuremath{#1})}
\newcommand{\accepted}[1]{accepted(\ensuremath{#1})}
\newcommand{\executed}[1]{executed(\ensuremath{#1})}
\newcommand{\sgexecuted}[1]{sg-executed(\ensuremath{#1})}
\newcommand{\execute}[1]{execute(\ensuremath{#1})}
\newcommand{\committed}[1]{sg-committed(\ensuremath{#1})}

\newcommand{\prepared}[1]{prepared(\ensuremath{#1})}

\newcommand{\ordered}[1]{committed(\ensuremath{#1})}
\newcommand{\prefix}[1]{prefix(\ensuremath{#1})}
\newcommand{\confirmview}[1]{startView(\ensuremath{#1})}
\newcommand{\checkpoint}[1]{checkpoint(\ensuremath{#1})}
\newcommand{\chkconfirm}[1]{chkConfirm(\ensuremath{#1})}

\newcommand{\inview}[1]{inView(\ensuremath{#1})}

\newcommand{\cref}[2]{Algorithm~\ref{#1} \emph{lines}:\{#2\}}

\begin{figure*}[ht!]
	\begin{center}
		\cbox{
			\small{
				$c,\req,\rep$ : Client $c$, request $\req$ from client and reply $\rep$ of $\req$.\\
				\clientDelivered{c,\req,\rep} - Client $c$ delivers response \rep for request \req.\\
				\before{\req,\req'} - Request \req is executed prior to request $\req'$, i.e., $\req'$ is executed based on execution of \req.\\
				$\vs_i$ : the set of replicas in synchronous group $i$.\\ 
				\accepted{c,\req,\rep,i} - Client $c$ receives $t+1$ matching replies of $req$ from every active replica in view $i$.\\
				\prefix{\req,\req',s_j} - Request $\req'$ is executed after execution of request $\req$ at replica $s_j$.\\
				\ordered{\req,i,\sn,s_j} - Active replica $s_j\in\vs_i$ has received $f+1$ matching \msgtag{prepare} or \msgtag{commit} messages.\\
				\committed{\req,i,\sn} - $\forall$ benign active replica $s_j\in\vs_i$: \ordered{\req,i,\sn,s_j}.\\
				\executed{\req,i,\sn,s_j} - Active replica $s_j\in\vs_i$ has executed request \req at sequence number \sn in its state.\\
				\sgexecuted{\req,i,\sn} - $\forall$ benign active replica $s_j\in\vs_i$: \executed{\req,i,\sn,s_j}.\\
				\prepared{\req,i,sn,s_j} - Active replica $s_j\in\vs_i$ has received \msgtag{prepare} message at $sn$ for $\req$.
			}}
		\end{center}
		\caption{\PCBFT\ proof notation.}
		\label{fig:proof}
	\end{figure*}
	
	\subsection{Safety (Consistency)}
	\label{sec:safety}
	
	\begin{theorem}{(safety)}\label{the:safety}
		If \clientDelivered{c,\req,\rep}, \clientDelivered{c',\req',\rep'}, and $\req\neq\req'$, then either \before{\req,\req'} or \before{\req',\req}.
	\end{theorem}
	
	To prove the safety property, we start from Lemma~\ref{lem:view} which shows a useful relation between predicates \clientDelivered{} and \accepted{}.
	
	\begin{lemma}{(view exists)}\label{lem:view}
		\clientDelivered{c,\req,\rep} $\Leftrightarrow$ $\exists$ view $i$: \accepted{c,\req,\rep,i}.
	\end{lemma}
	\tlproofs{By common case protocol \cref{alg:f1}{26-29} and \cref{alg:f2}{22-25}, client $c$ delivers a reply only upon it receives $t+1$ matching \msgtag{reply} messages from all active replicas in the same view. Conversely, upon client $c$ receives $t+1$ matching \msgtag{reply} messages from active replicas in the same view, it delivers the reply.}

	\begin{lemma}{(reply is correct)}\label{lem:reply}
		If \accepted{c,\req,\rep,i}, then \rep is the reply of \req executed by correct replica.
	\end{lemma}
	\tlproof{}{
		\item $\exists s_j\in\vs_i$: $s_j$ is correct.
		\\\artproof{Assumption of at most $t$ faulty replicas and $|\vs_i|=t+1$.}
		\item Client $c$ expects matching replies from $t+1$ active replicas in $\vs_i$.
		\\\artproof{By common case protocol \cref{alg:f1}{26-29} and \cref{alg:f2}{22-25}.}
		\item Q.E.D.
		\\\artproof{By 1 and 2.}
	}
	
	By Lemma~\ref{lem:view} and Lemma~\ref{lem:reply}, we assume $\exists$ view $i$ for $\req$ and $\exists i'$ for $\req'$, then we instead prove :
	
	\begin{theorem}{(safety)}\label{lem:accepted}
		If \accepted{c,\req,\rep,i} and \accepted{c',\req',\rep',i'}, then \before{\req,\req'} or   \before{\req',\req}.
	\end{theorem}
	
	Now we introduce sequence number.
	
	\begin{lemma}{(sequence number exists)}\label{lem:snexists}
		If \accepted{c,\req,\rep,i}, then $\exists$ sequence number $\sn$: \sgexecuted{\req,i,\sn}.
	\end{lemma}
	\tlproof{}{
		\item Client $c$ accepts \rep in view $i$ as reply of \req upon: \\
		(1) $c$ receives $\MSGREPLY$ messages with matching $ts$, \rep, \sn and $i$; and,\\
		(2) $\MSGREPLY$ messages are attested by $t+1$ active replicas in $\vs_i$.
		\\\artproof{By common case protocol \cref{alg:f1}{26-29} and \cref{alg:f2}{22-25}.}
		\item Benign active replica $s_j\in\vs_i$ sends $\MSGREPLY$ message for \req only upon $\exists\sn$ : \executed{\req,i,\sn,s_j}.
		\\\artproof{By common case protocol \cref{alg:f1}{21-25} and \cref{alg:f2}{18-21}.}
		\item Q.E.D.
		\\\artproof{By 1 and 2.}
	}
	
	By Lemma~\ref{lem:snexists}, we assume $\exists$ sequence number \sn for \req and $\exists \sn'$ for $\req'$. Then we instead prove:
	
	\begin{theorem}{(safety)}
		\label{the:executedordered}
		If \sgexecuted{\req,i,\sn}, \sgexecuted{\req',i',\sn'} and $\sn<\sn'$, then $\forall$ benign active replica $s_{j'}\in\vs_{i'}$: \prefix{\req,\req',s_{j'}}.
	\end{theorem}
	
	Towards the proof of Theorem~\ref{the:executedordered}, 
	we first prove several lemmas below (from Lemma~\ref{lem:sgexecuted} to Lemma~\ref{lem:executeordered}).
	
	Lemma~\ref{lem:sgexecuted} proves that if a request is executed by a benign active replica, then that request has been committed by the same replica.
	
	\begin{lemma}{}\label{lem:sgexecuted}
		If \executed{\req,i,\sn,s_j}, then \ordered{\req,i,\sn,s_j}.
	\end{lemma}
	\tlproofs{By common case protocol \cref{alg:f1}{10-21} and \cref{alg:f2}{10-18}, every benign active replica first commits a request by receiving $t+1$ matching \msgtag{prepare} or \msgtag{commit} messages, then it executes the request based on committed order.}
	
	\begin{lemma}{(\ordered{} is unique)}\label{lem:committedunique}
		If \ordered{\req,i,\sn,s_j} and \ordered{\req',i,\sn,s_{j'}}, then $\req=\req'$.
	\end{lemma}
	\tlproof{Proved by contradiction.}{
		\item We assume $\exists$ requests $\req$ and $\req'$ : \ordered{\req,i,\sn,s_j}, \ordered{\req',i,\sn,s_{j'}} and $\req\neq\req'$.
		\\\artproof{Contradiction assumption.}
		\item $\exists $ correct active replica $s_k\in\vs_i$ : $s_k$ has sent \msgtag{prepare} or \msgtag{commit} message for both $\req$ and $\req'$ at $\sn$ (i.e., $s_k$ has executed common case protocol \cref{alg:f1}{8-9} or \cref{alg:f1}{15-17}, or \cref{alg:f2}{7-9} or \cref{alg:f2}{14-15}, for both $\req$ and $\req'$).
		\\\artproof{By $|\vs_i|=t+1$, $\exists s_k:$ $s_k$ is correct; then by 1, common case protocol \cref{alg:f1}{18-20} or \cref{alg:f2}{16-17}, and definition of \ordered{}.} 
		\item Q.E.D.
		\\\artproof{By 2 and 1.} 
	}
	
	Lemma~\ref{lem:committedordered} locates at the heart of \PCBFT safety proof, which is proved by induction. By Lemma~\ref{lem:committedordered} we show that, if request $\req$ is committed at $\sn$ by every (benign) active replica in the same view, and, if request $\req'$ is committed by any replica in the preceding view at $\sn$, then $\req=\req'$.
	
	\begin{lemma}{(\committed{} is durable)}\label{lem:committedordered}
		If \committed{\req,i,\sn}, then $\forall i'>i$ : if \ordered{\req',i',\sn,s_{j'}} then $\req=\req'$.
	\end{lemma}
	\tlproof{}{
		\item We assume $\forall i''$ and $s_{j''}$ : $i\leq i''<i'$ and $s_{j''}\in\vs_{i''}$, if \ordered{\req'',i'',\sn,s_{j''}} then $\req=\req''$.
		\\\artproof{Inductive Hypothesis.}
		\item $\forall$ benign replica $s_{j'}\in\vs_{i'}$ : $s_{j'}$ has been waiting for \msgtag{view-change} messages from $\forall s_k\in\Pi$ within $2\Delta$ time. 
		\\\artproof{By \ordered{\req',i',\sn,s_{j'}}, $s_{k'}$ has generated \msgtag{prepare} or \msgtag{commit} message at $sn$; by view change protocol \cref{alg:vc}{23,26,28}, a benign active replica generates a \msgtag{prepare} or \msgtag{commit} message in view $i'$ only upon the replica has executed \cref{alg:vc}{16} in view $i'$; then by \cref{alg:vc}{13-15}.}
		\item $\exists s_{j'}\in\vs_{i'}$ : $s_{j'}$ is correct.
		\\\artproof{By $|\vs_{i'}|=t+1$ and at most $t$ faulty replicas.}
		\item During view change to $i'$, $s_{j'}$ has collected \msgtag{view-change} message $m$ from a correct active replica $s_j\in\vs_i$.
		\\\artproof{By 2 and 3, view change protocol \cref{alg:vc}{13-15} have been executed at $s_{j'}$; $s_{j'}$ polls all replicas for \msgtag{view-change} messages and waits for response from $t+1$ replicas as well as the timer set to $2\timeout$ to expire. Assume that $s_{j'}$ has received \msgtag{view-change} messages from $r\geq 1$ replicas in view $i$. The other $t+1-r$ replicas in view $i$ are either faulty or partitioned based on definitions. Among $r$ replicas which have replied, at most $t-(t+1-r)=r-1$ are faulty. Hence, at least one replica, say, $s_j\in\vs_i$ is correct and has replied with $m$.		
		}
		\item $m$ contains $t+1$ matching \msgtag{prepare} or \msgtag{commit} messages for request $\req''$ at sequence number $\sn$, generated in view $i''\geq i$.
		\\\artproof{By \cref{alg:vc}{6-7}, benign replicas process messages in ascending view order, so that commit log at $sn$ generated in view $i$ will not be replaced by any commit log generated in view $i'''<i$; then by 4 and \committed{\req,i,\sn}.}
		\item In view $i'$, $\forall s_{k'}\in\vs_{i'}:$ $s_{k'}$ can commit $\req''$, or any $\req'''$ which is committed in view $i'''>i''$ at $sn$.
		\\\artproof{By \cref{alg:vc}{19} and 5.}
		\item $\req''=\req'''=\req$.
		\\\artproof{By 4 and 5, $\req''$ is committed in $i''$ and $\req'''$ is committed in $i'''$, where $i'''>i''\geq i$; then by 1.}
		\item $\req'=\req$.
		\\\artproof{By 6, 7 and \ordered{\req',i',\sn,s_{j'}}.}
	}
	
	By Lemma~\ref{lem:committedordered} we can easily get Lemma~\ref{lem:committeddurable}.
	
	\begin{lemma}{}\label{lem:committeddurable}
		If \committed{\req,i,\sn} and \committed{\req',i',\sn}, then $\req=\req'$.
	\end{lemma}
	\tlproofs{By Lemma~\ref{lem:committedordered} and definition of \committed{}.}

	\begin{lemma}{}\label{lem:commitorder}
		If \executed{\req,i,\sn,s_j}, then $\forall \sn'<\sn: \exists \req' $ s.t. \ordered{\req', i,\sn',s_j}.
	\end{lemma}
	\tlproofs{By common case protocol \cref{alg:f1}{21-22} and \cref{alg:f2}{18-19}, correct active replicas execute requests based on order defined by committed sequence number; by \executed{\req,i,\sn,s_j} and $\sn'<\sn$, \executed{\req',i,\sn',s_j}; and, by Lemma~\ref{lem:sgexecuted}.}
	
	\begin{lemma}{(\executed{} in order)}\label{lem:lexecuteordered}
		If \ordered{\req,i,\sn,s_j}, \executed{\req',i,\sn',s_j} and $\sn<\sn'$, then \prefix{\req,\req',s_j}.
	\end{lemma}
	\tlproofs{By Lemma~\ref{lem:commitorder}, $\exists \req'' $ s.t. $\ordered{\req'',i,\sn,s_j}$; by Lemma~\ref{lem:committedunique}, $\req''=\req$; by common case protocol \cref{alg:f1}{21-22} and \cref{alg:f2}{18-19}, benign active replicas execute requests based on order defined by committed sequence number $\sn$ and $\sn'$; and, by $\sn<\sn'$.}
	
	\begin{lemma}{}\label{lem:executeordered}
		If \committed{\req,i,\sn}, \sgexecuted{\req',i,\sn'} and $\sn<\sn'$, then $\forall$ benign active replica $s_j$ : \prefix{\req,\req',s_j}.
	\end{lemma}
	\tlproofs{By Lemma~\ref{lem:lexecuteordered}.}
	
	Now we can prove Theorem~\ref{the:executedordered}.
	
	\tlproof{}{
		\item \committed{\req,i,\sn} and \committed{\req',i',\sn'}.
		\\\artproof{By \sgexecuted{\req,i,\sn}, \sgexecuted{\req',i',\sn'} and Lemma~\ref{lem:sgexecuted}.}
		\\\\\noindent When $i<i'$ : 
		\item \committed{\req,i',\sn}.
		\\\artproof{By \sgexecuted{\req',i',\sn'}, Lemma~\ref{lem:commitorder} and $\sn<\sn'$, $\exists \req'' : \committed{\req'',i',\sn}$; then by Lemma~\ref{lem:committeddurable}, \committed{\req,i,\sn} and $i<i'$, $\req''=\req$.}
		\item $\forall$ benign active replica $s_{j'}\in\vs_{i'}$: \prefix{\req,\req',s_{j'}}.
		\\\artproof{By \sgexecuted{\req',i',\sn'}, 2, $\sn<\sn'$ and Lemma~\ref{lem:executeordered}.}
		\\\\\indent When $i=i'$ : 
		\item $\forall$ benign active replica $s_j\in\vs_i$: \prefix{\req,\req',s_j}.
		\\\artproof{By 1, \sgexecuted{\req',i',\sn'}, $i=i'$, $\sn<\sn'$ and Lemma~\ref{lem:executeordered}.}
		\\\\\noindent When $i>i'$ : 
		\item $\exists \req''$ : \committed{\req'',i',\sn}.
		\\\artproof{By Lemma~\ref{lem:commitorder}, \sgexecuted{\req',i',\sn'} and $\sn<\sn'$.}
		\item $\req''=\req$.
		\\\artproof{By 5 and Lemma~\ref{lem:committeddurable}.} 
		\item $\forall$ benign active replica $s_{j'}\in\vs_{i'}$: \prefix{\req,\req',s_j}.
		\\\artproof{By 5, 6, \sgexecuted{\req',i',\sn'}, $\sn<\sn'$ and Lemma~\ref{lem:executeordered}.}
		\item Q.E.D.
		\\\artproof{By 3, 4 and 7.}
	}
	
	\subsection{Liveness (Availability)}\label{sec:prliveness}
	
	Before proving liveness property, we first prove two Lemmas (\ref{lem:livenessamongviews} and \ref{lem:livenessinaview}).
	
	\begin{lemma}\label{lem:livenessamongviews}
		If a correct client $c$ issues a request \req in view $i$, then eventually, either (1) \accepted{c,\req,\rep,i} or (2) \PCBFT changes view to $i+1$.
	\end{lemma}
	\tlproof{}{
		\item We assume \accepted{c,\req,\rep,i} is false, then we prove that eventually view $i$ is changed to $i+1$.
		\\\artproof{Equivalent.}
		\item Client $c$ sends $\req$ to every active replica upon $\timer_{c}$ expires.
		\\\artproof{By 1, $c$ is correct, and \cref{alg:liveness}{1-2}.}
		\item No replica in $\vs_i$ sent matching \msgtag{signed-reply} message for $\req$ to client $c$.
		\\\artproof{By 1, $c$ is correct and \cref{alg:liveness}{18-22}.}
		\item $\exists$ active replica $s_j\in\vs_i$: $s_j$ is correct.
		\\\artproof{By assumption $|\vs_i|=t+1$ and at most $t$ faulty replicas.}
		\item $s_j$ has not received $t+1$ matching signed \msgtag{reply} messages for $\req$.
		\\\artproof{By 3, 4 and \cref{alg:liveness}{18-22}.}
		\\\\\noindent Either,
		\item $s_j$ starts $timer_{req_c}$.
		\\\artproof{By 2, 4 and \cref{alg:liveness}{3,6}.}
		\item $s_j$ suspects view $i$ when $timer_{req_c}$ expires.
		\\\artproof{By 4, 5, 6 and \cref{alg:liveness}{8-10}.}
		\\\\\noindent or,
		\item $s_j$ starts $timer^{vc}_{i}$ in view change to $i$.
		\\\artproof{By \cref{alg:vc}{15}.}
		\item $s_j$ suspects view $i$ when $timer^{vc}_{i}$ expires.
		\\\artproof{By 2, 8 and \cref{alg:vc}{34-35}.}
		\item Q.E.D.
		\\\artproof{By 1 and 7, 9.}
	}
	
	\begin{lemma}\label{lem:livenessinaview}
		If a correct client $c$ issues a request \req in view $i$, the system is synchronous for a sufficient time and $\forall$ active replica $s_j\in\vs_i$: $s_j$ is correct, then eventually \accepted{c,\req,\rep,i}.
	\end{lemma}
	\tlproof{}{
		\item All active replicas in $\vs_i$ and $c$ follows protocol correctly.
		\\\artproof{$c$ is correct and $\forall$ active replica $s_j\in\vs_i$: $s_j$ is correct.}
		\item No timer expires.
		\\\artproof{By 1 and the system is synchronous.}
		\item No view change happens.
		\\\artproof{By 1 and \cref{alg:vc}{1-7}, no faulty replica in $\vs_i$, and no faulty passive replica in view $i$ can suspect view $i$ deliberately; and by 2, no correct replica in $\vs_i$ suspects view $i$.} 
		\item \accepted{c,\req,\rep,i}.
		\\\artproof{By 3 and Lemma~\ref{lem:livenessamongviews}.}
	}
	
	\begin{theorem}{(liveness)}\label{the:liveness}
		If a correct client $c$ issues a request \req, then eventually, \clientDelivered{c,\req,\rep}.
	\end{theorem}
	\tlproof{Proved by Contradiction.}{
		\item We assume \clientDelivered{c,\req,\rep} is always false.
		\\\artproof{Contradiction assumption.}
		\item If current view is $i$, then view is eventually changed to $i+1$.
		\\\artproof{By 1, Lemma~\ref{lem:view} and Lemma~\ref{lem:livenessamongviews}.}
		\item View change is executed for infinite times.
		\\\artproof{By 1 and 2, \cref{alg:liveness}{11-15} and \cref{alg:liveness}{1-2}, correct client $c$ always multicasts \msgtag{suspect} message and $req$ to every active replica in new view.}
		\item Eventually the system is synchronous.
		\\\artproof{Eventual synchrony assumption.}
		\item $\exists$ view $i'$: $\forall$ active replica $s_{j'}\in\vs_{i'}$ s.t. $s_{j'}$ is correct.
		\\\artproof{View change protocol is rounded among combinations of $2t+1$ replicas, among which there exists one synchronous group containing only correct active replicas.}
		\item \accepted{c,\req,\rep,i'}.
		\\\artproof{By 3, 4, 5, Lemma~\ref{lem:livenessinaview} and $c$ is correct.}
		\item Q.E.D.
		\\\artproof{By 1, 6, Lemma~\ref{lem:view} and contradiction.}
	}
	
	\subsection{Fault detection (FD)}\label{sec:prfaultdetection}
	
	In this section we prove that the fault detection mechanism is strong completeness and strong accuracy outside anarchy.

At first, in Definition~\ref{def:mmsg} we define the type of messages which can possibly violate consistency in anarchy.
	
	\begin{definition}{(non-crash faulty message)}\label{def:mmsg}
		In view change to $i$, a \msgtag{view-change} message $m$ from replica $s_k$ is a non-crash faulty message if :
		\par\setlength{\parindent}{2ex}
		(i) $m$ is sent to a correct active replica $s_j\in\vs_i$;\par
		(ii) $\exists$ view $i'<i$ and request $\req$ : $\committed{\req,i',\sn}$; \par
		(iii) at least one of two properties below is satisfied :
		\par\setlength{\parindent}{4ex}
		(1) $s_k\in\vs_{i'}$ and in $m$ : $PrepareLog[sn]$ is generated in view $i''<i'$; or,\par
		(2) in $m$ : $PrepareLog[sn].req\neq\req$ and $PrepareLog[sn]$ is generated in view $i''\geq i'$; and,
		\par\setlength{\parindent}{2ex}
		(iv) $\nexists i'''$ ($i'''>i''$ and $i'''>i'$) and $s_{k'''}\in\vs_{i'''}$ : $\ordered{\req,i''',\sn,s_{k'''}}$.
	\end{definition}

%
	
	Then we can prove:
	
	\begin{lemma}{}\label{lem:hm}
		If a \msgtag{view-change} message $m$ is not a non-crash faulty message, then $m$ cannot violate consistency in anarchy. 
	\end{lemma}
	\tlproof{Proved by Contradiction.}{
		\item If Definition~\ref{def:mmsg} property ($i$) is not satisfied, then either $m$ is sent to a non-crash faulty replica, based on our model we have no assumption on non-crash faulty replicas, so $m$ should not affect the state of any correct replica; or $m$ is sent to a crashed or passive replica, which just stops processing or ignores $m$.
		
		\item If Definition~\ref{def:mmsg} property ($ii$) is not satisfied, then $\req$ has not been committed by some correct replica in $\vs_{i'}$, hence \accepted{c,\req,\rep,i'} is not true.
		
		\item If neither of Definition~\ref{def:mmsg} property ($iii$).(1) or (2) is satisfied, then either $s_k\in\vs_{i'}$ and $m$ contains prepare log of $\req$ at $sn$ generated in view $i''\geq i‘$, so by \cref{alg:vcfd}{11-21}, $m$ facilitates $req$ to be committed in view $i$; or, if $s_k\notin\vs_{i'}$, then either $PrepareLog[sn].req$ is generated in $i''<i'$, even if $PrepareLog[sn].req\neq req$, based on \cref{alg:vcfd}{13,14,18} $PrepareLog[sn].req$ cannot be selected in view change to $i$ if no (faulty) replica in $i'$ sends inconsistent message (e.g., a prepare log generated in view lower than $i'$ by $s_{k'}\in\vs_{i'}$), hence we consider in this case $s_k$ is harmless; or $i''\geq i‘$ and $PrepareLog[sn].req=req$, the argument is the same as before.
		\item if Definition~\ref{def:mmsg} property ($iv$) is not satisfied, then $\exists i'''$ ($i'''>i''$ and $i'''>i'$) and $s_{k'''}\in\vs_{i'''}$ : $\ordered{\req,i''',\sn,s_{k'''}}$. In this case, to modify $req$ committed at $sn$, at least one of (faulty) replicas in $\vs_{i'''}$ has to send a non-crash faulty message; otherwise, based on \cref{alg:vcfd}{11-21}, any non-crash faulty message generated in $i''$ will be ignored.

	}
	
	Finally, we prove fault detection property : strong completeness and strong accuracy. Roughly speaking, (strong completeness) if a message is a \emph{non-crash faulty} message, then the sender will be detected eventually; otherwise, (strong accuracy) if a replica is correct, then it will never be detected.
	
	\begin{theorem}{(strong completeness)}\label{the:faultdetection}
		If a replica $s_k$ fails arbitrarily outside anarchy, in a way that would cause inconsistency in anarchy, then \PCBFT\ FD detects $s_k$ as faulty (outside anarchy).
	\end{theorem}
	\tlproof{}{
		\item By Lemma~\ref{lem:hm}, it is equivalent to prove : in view change to $i$, if $m$ is a non-crash faulty message from replica $s_k$, then correct active replica $s_j\in\vs_i$ detects the fault of $s_k$.
		\item By Definition~\ref{def:mmsg} property ($ii$), every correct replica $s_{k'}\in\vs_{i'}$ has commit log of $req$ at $sn$ generated in view equal to or higher than $i'$. Assume that the highest view in which commit log of $req$ is generated is $i_0$ ($i'\leq i_0<i$) .
		\\\artproof{By Lemma~\ref{lem:committedordered}.}		
				
		If in 2 $i_0=i'$ :
		
		\item Correct active replica $s_j\in\vs_i$ should receive $m'$ which contains commit log of \req generated in view $i'$ from correct active replica $s_{k'}\in\vs_{i'}$.
		\\\artproof{By outside anarchy, 2, Definition~\ref{def:mmsg} and Lemma~\ref{lem:committedordered}.}
		\item If $m$ satisfies Definition~\ref{def:mmsg} property ($iii$).($1$), then $s_j$ detects the fault of $s_k$.
		\\\artproof{By Definition~\ref{def:mmsg} property ($iii$).($1$), prepare log of $\req$ is not included in $m$; then by 3 and \cref{alg:fd}{3}, the fault is detected.}
		\item If $m$ satisfies Definition~\ref{def:mmsg} property ($iii$).($2$), then $s_j$ detects the fault of $s_k$.
		\\\artproof{By Definition~\ref{def:mmsg} property ($iii$).($2$), the prepare log at sequence number $sn$ is generated in view $i''<i$, then by 3 and \cref{alg:fd}{6} the fault of $s_k$ is detected.}
		\item If $m$ satisfies Definition~\ref{def:mmsg} property ($iii$).($3$), then $s_j$ detects the fault of $s_k$.
		\\\artproof{If in Definition~\ref{def:mmsg} property ($iii$).($3$) $i''=i'$, then by 3 and \cref{alg:fd}{6} the fault of $s_k$ is detected; otherwise, if $i''>i'$, then based on Lemma~\ref{lem:committedordered} $req$ must be retrieved by every correct active replica in view $i''$; hence by outside anarchy and \cref{alg:fd}{9-14} the fault of $s_k$ is detected.}
		
		If in 2 $i_0>i'$:
		\item Every replica (correct or faulty) in view $i_0$ has retrieved and prepared $req$ in view equal to or higher than $i_0$.
		\\\artproof{By 2, $i'''>i'$ and \cref{alg:vc}{26,28}.}
		\item In order to modify request committed at $sn$ (i.e., $req$), at least one of (faulty) replicas, say $s_{k'''}$ (in $\vs_{i_0}$ or not), has to send an inconsistent prepare log generated in view $i_1\geq i_0$. Hence, $s_k$ in this case is harmless.
		\\\artproof{By 7, $n=2t+1$, $i_0>i'$ and \cref{alg:vcfd}{19}.}
		\item Correct active replica $s_j\in\vs_i$ should receive $m'$ which contains commit log of \req generated in view $i_2$ ($i'\leq i_2 \leq i_0\leq i_1<i$) from correct active replica $s_{k'}\in\vs_{i'}$.
		\\\artproof{By outside anarchy, 2, Definition~\ref{def:mmsg} and Lemma~\ref{lem:committedordered}.}
		\item If $i_2<i_1$, then the fault of $s_{k'''}$ is detected by \cref{alg:fd}{9-16}, which is similar to discussion in 6; if $i_2=i_1$, then the fault of $s_{k'''}$ is detected by \cref{alg:fd}{3,6}, which is similar to discussion in 4 or 5.
		\item Q.E.D.
		\\\artproof{By 3, 4, 5 and 6 and 10.}
	}
	
	\begin{theorem}{(Strong accuracy)}\label{the:faultdetection}
		If a replica $s_k$ is benign (i.e., behaves faithfully), then \PCBFT\ FD will never detect $s_k$ as faulty.
	\end{theorem}
	\tlproof{}{
		\item It is equivalent to prove : in view change to $i$, if $s_k$ is benign and $s_k$ sends a \msgtag{view-change} message $m$ to all active replicas in view $i$, then no active replica in $\vs_i$ can detect $s_k$ as faulty.
		\\\artproof{Equivalent.}
		\\\\\noindent $\forall$ request \req, view $i'<i$ and replica $s_{j'}$ s.t. $s_k,s_{j'}\in\vs_{i'}$ and \ordered{\req,i',sn,s_{j'}}:
		\item $m$ contains prepare log of $\req'$ at $sn$ generated in view $i''\geq i'$.
		\\\artproof{By common case protocol \cref{alg:f1}{9,17}, \cref{alg:f2}{9,15} and view-change \cref{alg:vcfd}{1}, $s_k$ sends a prepare log at sequence number $sn$ once $s_k$ prepared a request at $sn$; by \cref{alg:vc}{6-7}, correct replicas process messages in ascending view order, hence $i''\geq i'$.}
		\item $s_k$ will not be detected by \cref{alg:fd}{3} due to \ordered{\req,i',sn,s_{j'}}.
		\\\artproof{By 2 and \cref{alg:fd}{3}.}
		
		
		\item No other request $\req''\neq\req'$ is committed by any replica at sequence number $sn$ in view $i'$.
		\\\artproof{By $s_k$ is correct and Lemma~\ref{lem:committedunique}.}
		
		\item $s_k$ will not be detected by \cref{alg:fd}{6} due to \ordered{\req,i',sn,s_{j'}}.
		\\\artproof{By 4 and \cref{alg:fd}{6}.}
		\item $s_k$ will not be detected by \cref{alg:fd}{9} due to \ordered{\req,i',sn,s_{j'}}.
		\\\artproof{By $s_k$ is correct, $s_k$ did not generate or accept any incorrect prepare log during view-change to view $i''$; by \cref{alg:fd}{9}, \cref{alg:vcfd}{3-7} and Lemma~\ref{lem:committedordered}, no conflict $vcSet^{i''}_{k'}$ and $finalProof_{s_{k'}}[i'']$ exists in view $i''$ at any active replica.}
		
%
%
		
		\item Q.E.D.
		\\\\\artproof{By 3, 5 and 6.}
	}
	
	We can easily prove that if a fault is detected by any correct replica, then the fault is detected by every replica eventually.
	
	\begin{lemma}{}\label{the:reliable}
		In view change to $i$, if a correct active replica $s_j\in\vs_i$ detects the fault of $s_k$, then eventually every correct replica detects the fault of $s_k$.
	\end{lemma}
	\tlproofs{By \cref{alg:fd}{6-7}.}

\clearpage
\section{Reliability analysis (examples)}
In Table~\ref{tab:9_t1} and~\ref{tab:9_t2} we show the nines of consistency of each model when $t=1$ and $t=2$ for some practical values of $\ninesbenign$, $\ninessynchrony$ and $\ninescorrect$; in Table~\ref{tab:9a_t1} and~\ref{tab:9a_t2} we show the nines of availability of each model when $t=1$ and $t=2$ for some practical values of $\ninesnormal$ and $\ninesbenign$.

\begin{table*}[hbtp]
			\footnotesize
			\centering
			\begin{scriptsize}
				\begin{tabular}{|c|c?c|c|c|c|c|c?c|}
					\cline{3-8}
					\multicolumn{2}{c?}{} & \multicolumn{6}{c?}{$\ninesofC(\PCBFT_{t=1})$}\\
					\hline
					\multirow{2}{*}{$\ninesbenign$} & \multirow{2}{*}{$\ninesofC(CFT_{t=1})$} & \multirow{2}{*}{$\ninescorrect$} & \multicolumn{5}{c?}{$\ninessynchrony$} & \multirow{2}{*}{$\ninesofC(BFT_{t=1})$}\\
					\cline{4-8}
					 & & & \emph{2} & \emph{3} & \emph{4} & \emph{5} & \emph{6} &\\
					\thickhline
					\emph{3} & 2 & \emph{2} & 3 & 4 & 4 & 4 & 4 & 5\\
					\hline
					\multirow{2}{*}{\emph{4}} & \multirow{2}{*}{3} & \emph{2} & 4 & 5 & 5 & 5 & 5 & \multirow{2}{*}{7}\\
					\cline{3-8}
					& & \emph{3} & 5 & 5 & 6 & 6 & 6 &\\		
					\hline
					\multirow{3}{*}{\emph{5}} & \multirow{3}{*}{4} & \emph{2} & 5 & 6 & 6 & 6 & 6 & \multirow{3}{*}{9}\\
					\cline{3-8}
					& & \emph{3} & 6 & 6 & 7 & 7 & 7 &\\
					\cline{3-8}
					& & \emph{4} & 6 & 7 & 7 & 8 & 8 &\\
					\hline
					\multirow{4}{*}{\emph{6}} & \multirow{4}{*}{5} & \emph{2} & 6 & 7 & 7 & 7 & 7 & \multirow{4}{*}{11}\\
					\cline{3-8}
					& & \emph{3} & 7 & 7 & 8 & 8 & 8 &\\
					\cline{3-8}
					& & \emph{4} & 7 & 8 & 8 & 9 & 9 &\\
					\cline{3-8}
					& & \emph{5} & 7 & 8 & 9 & 9 & 10 &\\
					\hline
					\multirow{5}{*}{\emph{7}} & \multirow{5}{*}{6} & \emph{2} & 7 & 8 & 8 & 8 & 8 & \multirow{5}{*}{13}\\
					\cline{3-8}
					& & \emph{3} & 8 & 8 & 9 & 9 & 9 &\\
					\cline{3-8}
					& & \emph{4} & 8 & 9 & 9 & 10 & 10 &\\
					\cline{3-8}
					& & \emph{5} & 8 & 9 & 10 & 10 & 11 &\\
					\cline{3-8}
					& & \emph{6} & 8 & 9 & 10 & 11 & 11 &\\
					\hline
					\multirow{6}{*}{\emph{8}} & \multirow{6}{*}{7} & \emph{2} & 8 & 9 & 9 & 9 & 9 & \multirow{6}{*}{15}\\
					\cline{3-8}
					& & \emph{3} & 9 & 9 & 10 & 10 & 10 &\\
					\cline{3-8}
					& & \emph{4} & 9 & 10 & 10 & 11 & 11 &\\
					\cline{3-8}
					& & \emph{5} & 9 & 10 & 11 & 11 & 12 &\\
					\cline{3-8}
					& & \emph{6} & 9 & 10 & 11 & 12 & 12 &\\
					\cline{3-8}
					& & \emph{7} & 9 & 10 & 11 & 12 & 13 &\\
					\hline
				\end{tabular}
			\end{scriptsize}
		\caption{$\ninesofC(CFT_{t=1})$, $\ninesofC(\PCBFT_{t=1})$ and $\ninesofC(BFT_{t=1})$ values when $3\leq \ninesbenign\leq 8$, $2\leq\ninessynchrony\leq 6$ and $2\leq\ninescorrect<\ninesbenign$.}%
		\label{tab:9_t1}
\end{table*}

\begin{table*}[hbtp]
			\footnotesize
			\centering
			\begin{scriptsize}
				\begin{tabular}{|c|c?c|c|c|c|c|c?c|}
					\cline{3-8}
					\multicolumn{2}{c?}{} & \multicolumn{6}{c?}{$\ninesofC(\PCBFT_{t=2})$}\\
					\hline
					\multirow{2}{*}{$\ninesbenign$} & \multirow{2}{*}{$\ninesofC(CFT_{t=2})$} & \multirow{2}{*}{$\ninescorrect$} & \multicolumn{5}{c?}{$\ninessynchrony$} & \multirow{2}{*}{$\ninesofC(BFT_{t=2})$}\\
					\cline{4-8}
					 & & & \emph{2} & \emph{3} & \emph{4} & \emph{5} & \emph{6} &\\
					\thickhline
					\emph{3} & 2 & \emph{2} & 4 & 5 & 5 & 5 & 5 & 7\\
					\hline
					\multirow{2}{*}{\emph{4}} & \multirow{2}{*}{3} & \emph{2} & 5 & 6 & 6 & 6 & 6 & \multirow{2}{*}{10}\\
					\cline{3-8}
					& & \emph{3} & 6 & 7 & 8 & 8 & 8 &\\		
					\hline
					\multirow{3}{*}{\emph{5}} & \multirow{3}{*}{4} & \emph{2} & 6 & 7 & 7 & 7 & 7 & \multirow{3}{*}{13}\\
					\cline{3-8}
					& & \emph{3} & 7 & 8 & 9 & 9 & 9 &\\
					\cline{3-8}
					& & \emph{4} & 7 & 9 & 10 & 11 & 11 &\\
					\hline
					\multirow{4}{*}{\emph{6}} & \multirow{4}{*}{5} & \emph{2} & 7 & 8 & 8 & 8 & 8 & \multirow{4}{*}{16}\\
					\cline{3-8}
					& & \emph{3} & 8 & 9 & 10 & 10 & 10 &\\
					\cline{3-8}
					& & \emph{4} & 8 & 10 & 11 & 12 & 12 &\\
					\cline{3-8}
					& & \emph{5} & 8 & 10 & 12 & 13 & 14 &\\
					\hline
					\multirow{5}{*}{\emph{7}} & \multirow{6}{*}{6} & \emph{2} & 8 & 9 & 9 & 9 & 9 & \multirow{6}{*}{19}\\
					\cline{3-8}
					& & \emph{3} & 9 & 19 & 11 & 11 & 11 &\\
					\cline{3-8}
					& & \emph{4} & 9 & 11 & 12 & 13 & 13 &\\
					\cline{3-8}
					& & \emph{5} & 9 & 11 & 13 & 14 & 15 &\\
					\cline{3-8}
					& & \emph{6} & 9 & 11 & 13 & 15 & 16 &\\
					\hline
					\multirow{6}{*}{\emph{8}} & \multirow{6}{*}{7} & \emph{2} & 9 & 10 & 10 & 10 & 10 & \multirow{6}{*}{22}\\
					\cline{3-8}
					& & \emph{3} & 10 & 11 & 12 & 12 & 12 &\\
					\cline{3-8}
					& & \emph{4} & 10 & 12 & 13 & 14 & 14 &\\
					\cline{3-8}
					& & \emph{5} & 10 & 12 & 13 & 15 & 16 &\\
					\cline{3-8}
					& & \emph{6} & 10 & 12 & 14 & 16 & 17 &\\
					\cline{3-8}
					& & \emph{7} & 10 & 12 & 14 & 16 & 18 &\\
					\hline
				\end{tabular}
			\end{scriptsize}
		\caption{$\ninesofC(CFT_{t=2})$, $\ninesofC(\PCBFT_{t=2})$ and $\ninesofC(BFT_{t=2})$ values when $3\leq \ninesbenign\leq 8$, $2\leq\ninessynchrony\leq 6$ and $2\leq\ninescorrect<\ninesbenign$.}%
		\label{tab:9_t2}
\end{table*}

\begin{table*}[hbtp]
			\footnotesize
			\centering
			\begin{scriptsize}
				\begin{tabular}{|c?c|c|c|c|c|c?c|c|}
					\cline{2-7}
					\multicolumn{1}{c?}{} & \multicolumn{6}{c?}{$\ninesofA(CFT_{t=1})$}\\
					\hline
					\multirow{2}{*}{$\ninesnormal$} & \multicolumn{6}{c?}{$\ninesbenign$} & \multirow{2}{*}{$\ninesofA(BFT_{t=1})$} & \multirow{2}{*}{$\ninesofA(\PCBFT_{t=1})$}\\
					\cline{2-7}
					 & \emph{3} & \emph{4} & \emph{5} & \emph{6} & \emph{7} & \emph{8} & &\\
					\thickhline
					\emph{2} & 2 & 3 & 3 & 3 & 3 & 3 & 3 & 3\\
					\hline
					\emph{3} & \cellcolor{lightgray} & 3 & 4 & 5 & 5 & 5 & 5 & 5\\
					\hline
					\emph{4} & \cellcolor{lightgray} & \cellcolor{lightgray} & 4 & 5 & 6 & 7 & 7 & 7\\
					\hline
					\emph{5} & \cellcolor{lightgray} & \cellcolor{lightgray} & \cellcolor{lightgray} & 5 & 6 & 7 & 9 & 9\\
					\hline
					\emph{6} & \cellcolor{lightgray} & \cellcolor{lightgray} & \cellcolor{lightgray} & \cellcolor{lightgray} & 6 & 7 & 11 & 11\\
					\hline
				\end{tabular}
			\end{scriptsize}
		\caption{$\ninesofA(CFT_{t=1})$, $\ninesofA(BFT_{t=1})$ and $\ninesofA(\PCBFT_{t=1})$ values when $2\leq \ninesnormal\leq 6$ and $\ninesnormal<\ninesbenign\leq 8$.}%
		\label{tab:9a_t1}
\end{table*}

\begin{table*}[hbtp]
			\footnotesize
			\centering
			\begin{scriptsize}
				\begin{tabular}{|c?c|c|c|c|c|c?c|c|}
					\cline{2-7}
					\multicolumn{1}{c?}{} & \multicolumn{6}{c?}{$\ninesofA(CFT_{t=2})$}\\
					\hline
					\multirow{2}{*}{$\ninesnormal$} & \multicolumn{6}{c?}{$\ninesbenign$} & \multirow{2}{*}{$\ninesofA(BFT_{t=2})$} & \multirow{2}{*}{$\ninesofA(\PCBFT_{t=2})$}\\
					\cline{2-7}
					 & \emph{3} & \emph{4} & \emph{5} & \emph{6} & \emph{7} & \emph{8} & &\\
					\thickhline
					\emph{2} & 2 & 3 & 4 & 4 & 4 & 5 & 4 & 5\\
					\hline
					\emph{3} & \cellcolor{lightgray} & 3 & 4 & 5 & 6 & 7 & 7 & 8\\
					\hline
					\emph{4} & \cellcolor{lightgray} & \cellcolor{lightgray} & 4 & 5 & 6 & 7 & 10 & 11\\
					\hline
					\emph{5} & \cellcolor{lightgray} & \cellcolor{lightgray} & \cellcolor{lightgray} & 5 & 6 & 7 & 13 & 14\\
					\hline
					\emph{6} & \cellcolor{lightgray} & \cellcolor{lightgray} & \cellcolor{lightgray} & \cellcolor{lightgray} & 6 & 7 & 16 & 17\\
					\hline
				\end{tabular}
			\end{scriptsize}
		\caption{$\ninesofA(CFT_{t=2})$, $\ninesofA(BFT_{t=2})$ and $\ninesofA(\PCBFT_{t=2})$ values when $2\leq \ninesnormal\leq 6$ and $\ninesnormal<\ninesbenign\leq 8$.}%
		\label{tab:9a_t2}
\end{table*}

\label{app:reliability}

\end{appendices}
	
\end{document}